\documentclass[10pt,letterpaper]{article}
\usepackage{opex3}
\usepackage[draft]{hyperref}
\usepackage{amsmath}
\usepackage{amssymb}
\usepackage{bm}
\usepackage{graphicx}
\usepackage{subfigure}
\usepackage{float}
\def\ep{\varepsilon}
\def\dfrac{\displaystyle\frac}

\begin{document}

\title{Finite frequency external cloaking with complementary bianisotropic media}

\author{Yan Liu,$^{1,^*}$ Boris Gralak,$^1$ Ross. C. McPhedran,$^{2}$ and Sebastien Guenneau$^1$}

\address{$^1$ Ecole Centrale Marseille, CNRS, Aix-Marseille Universit\'e, Institut Fresnel\\
Campus de Saint-J\'er\^ome, 13013 Marseille, France \\
$^2$ School of Physics, The University of Sydney, Sydney, NSW 2006, Australia}

\email{$^*$yan.liu@fresnel.fr} 



\begin{abstract}
We investigate the twofold functionality of a cylindrical shell consisting of a negatively refracting heterogeneous bianisotropic (NRHB) medium deduced from geometric transforms. The numerical simulations indicate that the shell enhances their scattering by a perfect electric conducting (PEC) core, whereas it considerably reduces the scattering of electromagnetic waves by closely located dipoles when the shell surrounds a bianisotropic core. The former can be attributed to a homeopathic effect, whereby a small PEC object scatters like a large one as confirmed by numerics, while the latter can be attributed to space cancelation of complementary bianisotropic media underpinning anomalous resonances counteracting the field emitted by small objects (external cloaking). Space cancellation is further used to cloak a NRHB finite size object located nearby a slab of NRHB with a hole of same shape and opposite refracting index. Such a finite frequency external cloaking is also achieved with a NRHB cylindrical lens. Finally, we investigate an ostrich effect whereby the scattering of NRHB slab and cylindrical lenses with simplified parameters hide the presence of dipoles in the quasi-static limit.
\end{abstract}


\bibliographystyle{osajnl}
\bibliography{Bib_oe}
\section{Introduction}
In the past seven years, there has been a surge of interest in electromagnetic (EM) metamaterials deduced from the coordinate transformation approach proposed by Leonhardt \cite{Leonhardt06} and Pendry \cite{Pendry06}, such as invisibility cloaks designed through the blowup of a point \cite{Leonhardt06,Pendry06}, or space folding \cite{Chen10,Kadic11,Milton05,Milton06,Milton08} --the latter being based upon the powerful concept of complementary media introduced by Pendry and Ramakrishna ten years ago \cite{Pendry2003} -- or even superscatterers \cite{Yang08}.
Transformation optics is a useful mathematical tool enabling a deeper analytical insight into the scattering properties of EM fields in metamaterials. Geometric transforms can be chosen properly to design the metamaterials.

In this work, we make use of complementary media \cite{Pendry2003} and geometric transforms in order to design a heterogeneous bianisotropic shell behaving either as a superscatterer (SS) or an external cloak, depending upon whether its core is a perfect electric conductor (PEC) or certain bianisotropic medium. Indeed, the twofold functionality of the bianisotropic cylindrical shell which we propose displays the similar homeopathic effect to the dielectric shell studied in \cite{Nicorovici94}. Surface plasmon type resonances are visible on its interfaces when a set of dipoles appears to be in its close neighbourhood, and this leads to the similar cloaking to \cite{Milton06}, although in the present case this cloaking occurs at any frequency. It is also interesting to make large objects of negatively refracting media invisible when they are located closeby a slab or a cylindrical lens of opposite permittivity, permeability and magneto-coupling parameters with a hole of same shape as the object to hide. All these aforementioned bianisotropic cloaks work at finite frequency, but it is interesting to simply their optical parameters and check to which extent invisibility is preserved: We argue one can achieve an ostrich effect \cite{Nicorovici08} whereby it is virtually impossible to detect the scattering of set of dipoles located nearby a (much visible) slab or cylindrical perfect lens at quasi-static frequencies.

\section{SS through complementary bianisotropic media}
The source-free Maxwell-Tellegen's equations in a bianisotropic medium can be expressed as
\begin{equation}
\begin{array}{ll}
\nabla \times {\bf E} &= \omega {\underline{\underline{\xi}}} {\bf E}+i \omega {\underline{\underline{\mu}}} {\bf H}\\[2mm]
\nabla \times {\bf H} &= -i \omega {\underline{\underline{\ep}}} {\bf E}+\omega {\underline{\underline{\xi}}} {\bf H}\\
\end{array}
\label{maxtel}
\end{equation}
with $\omega$ the wave frequency, ${\underline{\underline{\ep}}}$ the permittivity, ${\underline{\underline{\mu}}}$ the permeability and ${\underline{\underline{\xi}}}$ the tensor of magneto-electric coupling. These equations retain their form under geometric changes \cite{Novitsky12,yanl13}, which can be derived as an extension of Ward and Pendry's result \cite{Ward96}.

Considering a map $\phi: {\bf x} \rightarrow {\bf x}' \,({\bf x}, {\bf x}' \in \mathbb{R}^3)$ described by ${\bf x}'({\bf x})$ (i.e. ${\bf x}'$ is given as a function of ${\bf x}$), the electromagnetic fields in the two coordinate systems satisfy ${\bf E}({\bf x})={\bf J}^{-{\rm T}}{\bf E'}({\bf x}')$, ${\bf H}({\bf x})={\bf J}^{-{\rm T}}{\bf H'}({\bf x}')$ \cite{Zolla07,Novitsky12}, and the parameter tensors satisfy
\begin{equation}
  {\underline{\underline{v}}}'= {\bf J}^{-1} {\underline{\underline{v}}} {\bf J}^{-{\rm T}} \det({\bf J}) , \quad v=\ep,\mu,\xi
\label{trans}
\end{equation}
where ${\bf J}$ is the Jacobian matrix of the coordinate transformation: ${\bf J}={\partial {\bf x}}/{\partial {\bf x'}}$, and the inverse ${\bf J}^{-1}={\partial {\bf x'}}/{\partial {\bf x}}$. Moreover, ${\bf J}^{-\rm T}$ denotes the inverse transposed Jacobian.

\begin{figure}[!htb]
    \centering
    \includegraphics[width=0.7\textwidth]{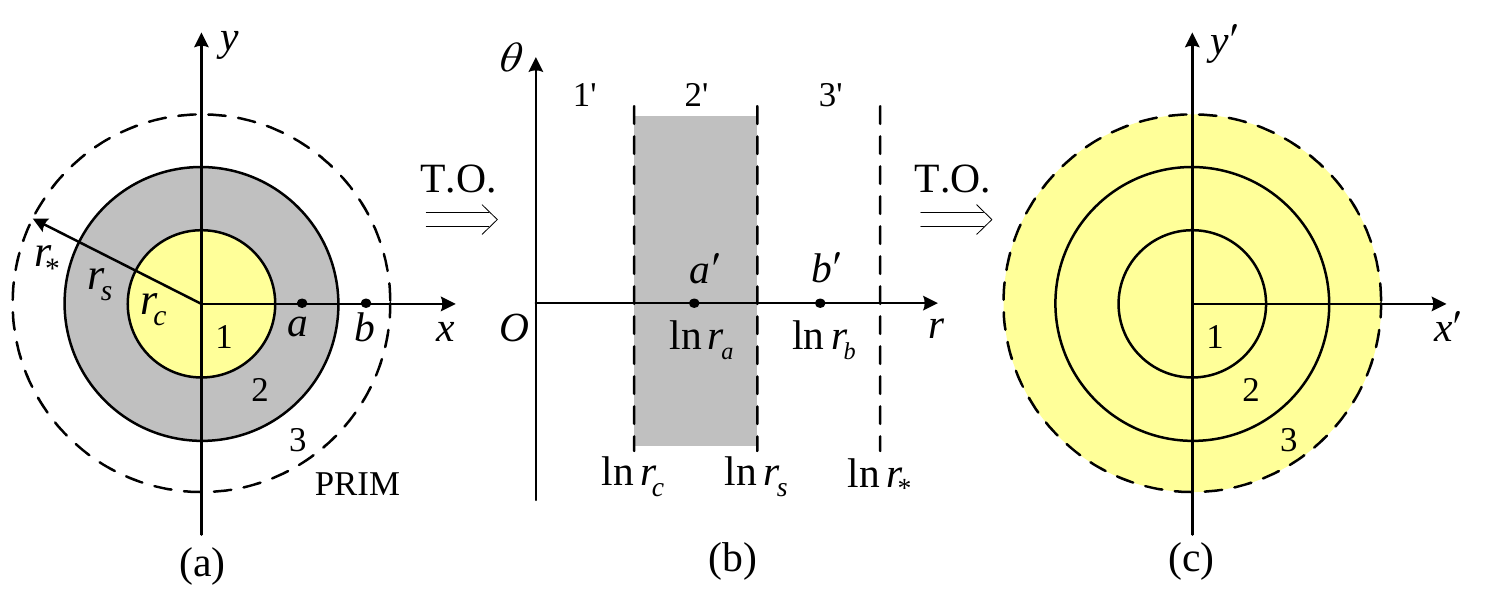}
    \caption{(a) Schematic of cylindrical lens consisting of a negatively refracting bianisotropic ring (grey color) $r_c<r\leq r_s$ which optically cancels out the positively refracting ring $r_s<r\leq r_*$ (white color). (b) Mapping of the three axisymmetric regions of panel (a) into layers in the new coordinate system $(r,\theta,z)$. (c) The enlarged region 1 in final Cartesian coordinates.}
    \label{figure1}
\end{figure}

Let us consider a cylindrical lens consisting of three regions: a core ($r\leq r_c$), a shell ($r_c<r\leq r_s$) and a matrix ($r>r_s$), which are filled with bianisotropic media as shown in Fig. \ref{figure1}(a) in Cartesian coordinates. The parameter in region $p$ ($p=1,2,3$) is denoted by ${\underline{\underline{v}}}^{(p)}=v_0{\bf diag} (v^{(p)}_x,v^{(p)}_y,v^{(p)}_z)$ ($v=\varepsilon,\mu,\xi$), region 3 is isotropic:  ${\underline{\underline{v}}}^{(3)}=v_0 I$ with $I$ the identity matrix.

To design a SS with an enhanced optical scattering cross section, i.e. the region 1 appears to be optically enlarged up to the boundary of region 3, the sum of optical paths 2+3 should be zero, which can be achieved by a pair of complementary bianisotropic media \cite{yanl13} according to the generalized lens theorem \cite{Pendry2003}.

Firstly, we introduce a map from Cartesian to cylindrical coordinates $(r,\theta,z)$ defined by \cite{Pendry2003,Guenneau05}
\begin{equation}
x=\exp(r)\cos\theta,\quad y=\exp(r)\sin\theta, \quad z=z
\label{cdtrans}
\end{equation}
and the Jacobian matrix is
\begin{equation}
  {\bf J}_{xr}=\dfrac{\partial(x,y,z)}{\partial(r,\theta,z)}
  =\left[\begin{array}{ccc}
    \exp(r)\cos\theta & -\exp(r)\sin\theta & 0\\
    \exp(r)\sin\theta & \exp(r)\cos\theta & 0 \\
    0 & 0 & 1
  \end{array}\right]
\end{equation}
Region $p$ is mapped to region $p'$ in the new coordinates. The transformed tensors of region 3' can be derived from (\ref{trans}) as ${\underline{\underline{v}}}'^{(3)}=v_0 {\bf diag}(1,1,\exp(2r))$. Furthermore, according to the generalized perfect lens theorem, the region 2' should be designed as the complementary medium of region 3', i.e. region 2'(respectively 3') is mirror imaged onto region 3'(respectively 2') along the axis $r=\ln r_s$, see Fig.\ref{figure1}(b). More precisely, we have
\begin{equation}
  {\underline{\underline{v}}}'^{(2)}(a')=-{\underline{\underline{v}}}'^{(3)}(b')= -v_0 {\bf diag}\big(1,1,\exp{(2 \ln r_b)}\big) \,,
\end{equation}
where $ r_b=r_s^2/r_a$ is derived from $\ln r_b- \ln r_s =\ln r_s- \ln r_a$. Moreover, the boundary $r_*$ of region 3 can also be fixed by $\ln r_*- \ln r_s =\ln r_s- \ln r_c$. Finally, we go back to the Cartesian coordinates through an inverse transformation with ${\bf J}_{xr}={\bf J}^{-1}_{rx}$, and we obtain
\begin{equation}
{\underline{\underline{v}}}^{(2)}(a)=-v_0 {\bf diag}\big(1,1,\exp{(2 \ln r_b)} \exp{(-2\ln r_a})\big)=-v_0 {\bf diag}(1,1,r_s^4/r_a^4).
\end{equation}
Regarding the tensors ${\underline{\underline{v}}}^{(1)}$ in region 1, if we define a function $F({\bf x})$ which enlarges the region 1 to fill up the optically canceled region (regions 2+3) as shown in Fig. \ref{figure1}(c), and suppose the new parameter of the enlarged region is ${\underline{\underline{v}}}_{\rm eff}$, then we can fix ${\underline{\underline{v}}}^{(1)}$ from the reverse of (\ref{trans}). If the scaling factor is $\gamma$ in the $x$-, $y$-directions while it is equal to 1 in $z$-direction for $r\leq r_c$, i.e. $(x',y',z')=F(x,y,z)=(\gamma x,\gamma y,z)$, then we have ${\underline{\underline{v}}}^{(1)}={\underline{\underline{v}}}_{\rm eff} {\bf diag}(1,1,\gamma^2)$. On the other hand, the boundary $r_c$ of region 1 is enlarged to $r_*$ as discussed above, hence $\gamma=r_*/r_c=r_s^2/r_c^2$.

Considering a transparent SS with ${\underline{\underline{v}}}_{\rm eff}/v_0= I$, then the relative permittivity, permeability and magneto-electric tensors of those three regions are
\begin{equation}
\begin{array}{cccc}
  v_x^{(1)}=+1, & v_y^{(1)}=+1, & v_z^{(1)}=+r_s^4/r_c^4, & r\leq r_c \\
  v_x^{(2)}=-1, & v_y^{(2)}=-1, & v_z^{(2)}=-r_s^4/r^4, & r_c < r\leq r_s \\
  v_x^{(3)}=+1, & v_y^{(3)}=+1, & v_z^{(3)}=+1, & r_s < r
\end{array}
\label{para}
\end{equation}

\begin{figure}[!htb]
    \centering
    \includegraphics[width=0.7\textwidth]{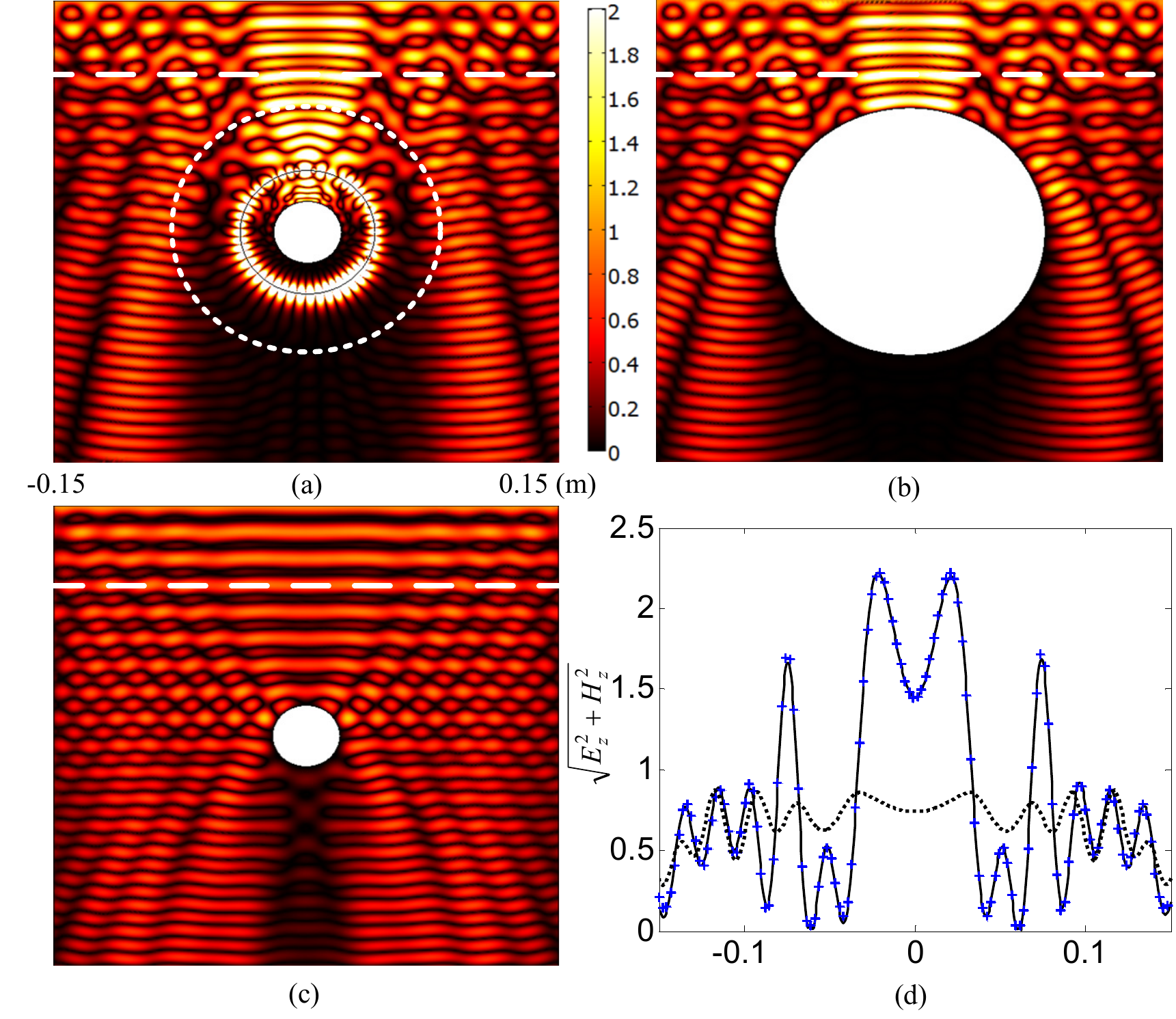}
    \caption{Plots of $\sqrt{E_z^2+H_z^2}$ under a TE polarized incidence with frequency $8.7$GHz: (a) Cylindrical lens made of bianisotropic media with parameters as in (\ref{para}), $r_s=0.04$m (shell radius) and a PEC boundary at $r=r_c=0.02$m; (b) Enlarged PEC boundary at $r_*=r^2_s/r_c=0.08$m in bianisotropic matrix; (c) Core with PEC boundary at $r=r_c$ in bianisotropic matrix; (d) Comparison of the plots of $\sqrt{E_z^2+H_z^2}$ on the intercepting line for (a) (solid curve), (b)(crosses) and (c) (dotted curve).}
    \label{figure2}
\end{figure}

Numerical illustration is carried out with COMSOL Multiphysics, the finite element method (FEM) result of a PEC core surrounded by a cylindrical shell consisting of bianisotropic media is shown in Fig. \ref{figure2}(a), while the equivalent PEC cylinder with $r_*$ is shown for comparison in Fig. \ref{figure2}(b), a TE polarized ($E_z$ is perpendicular to the $x$-$y$ plane) plane wave with frequency $8.7$GHz is incident from above. In these three regions, $v_0=\ep_0,\, \mu_0$ are the permittivity, permeability of the vacuum, while $\xi_0=0.99/c_0$ with $c_0$ the velocity of light in vacuum to ensure convergence of the numerical algorithm; and the radii are $r_c=0.02$m, $r_s=0.04$m. It can be seen that the scattered fields in Fig. \ref{figure2}(a) and (b) are quite similar outside the disc of radius $r_*=r^2_s/r_c$ (equivalent for an external observer to a disc of radius $r_*$ shown in Fig. \ref{figure2}). Moreover, the scattering by a PEC core is shown for comparison in panel (c). The profiles of the magnitude $\sqrt{E_z^2+H_z^2}$ of the scattered fields along the white dotted line depicted in panels (a)-(c) of Fig. \ref{figure2} are drawn in panel (d) with solid, crosses and dotted curves, respectively. The solid curve and crosses are nearly superimposed, unlike for the dotted curve, which proves the super scattering effect for a cylindrical lens with complementary bianisotropic media, similarly to the achiral case \cite{Yang08}.

\section{Bianisotropic SS through the space folding technique}
To understand how the SS works, we introduce a space folding technique. We consider the geometric transform which includes first a map from the Cartesian system to cylindrical coordinates through $x=r\cos\theta$, $y=r\sin\theta$, $z=z$.  Then a stretched cylindrical coordinates $(r',\theta',z')$ is introduced through a radial transform $r'=f(r)$ while $\theta'=\theta$ and $z'=z$. Finally we go back to Cartesian coordinates $(x',y',z')$. This compound transform leads to a Jacobian matrix
\begin{equation}
\begin{array}{l}
  {\bf J}_{xx'}={\bf J}_{xr}{\bf J}_{rr'}{\bf J}_{r'x'}={\bf R}(\theta){\bf diag}(1,r,1){\bf diag}\left(g',1,1\right) {\bf diag}(1,\dfrac{1}{r'},1){\bf R}^{-1}(\theta') \\
  \quad\quad={\bf R}(\theta) {\bf diag} (g',\dfrac{g}{r'},1) {\bf R}(-\theta')
\end{array}
\end{equation}
where $g(r')=r$ is the inverse function of $f$, and
\begin{equation}
  {\bf R}(\theta)=\left[\begin{array}{ccc}
    \cos(\theta) & -\sin(\theta) & 0 \\
    \sin(\theta) & \cos(\theta) & 0 \\
    0 & 0 & 1
  \end{array}\right]
\end{equation}
The transformation matrix (representation of metric tensor) reads
\begin{equation}
  {\bf T}^{-1}_{xx'}=\big({\bf J}^{\rm T}_{xx'}{\bf J}_{xx'}/{\rm det}({\bf J}_{xx'})\big)^{-1} ={\bf R}(\theta') {\bf diag}(\dfrac{g}{g'r'},\dfrac{g'r'}{g},\dfrac{g'g}{r'}) {\bf R}(-\theta')
  \label{cartpol}
\end{equation}
If the parameter ${\underline{\underline{v}}}$ in the coordinates $(x,y,z)$ is isotropic, then the transformed parameter in new coordinates $(x',y',z')$ is ${\underline{\underline{v}}}'= {\underline{\underline{v}}}{\bf T}^{-1}_{xx'}$ according to (\ref{trans}) \cite{Zolla07}.

For any such bianisotropic media with a translational invariance along the $z'=z$ axis, we can write the electromagnetic field in stretched cylindrical coordinates, wherein ${\bf E} = \left(E_{r'},E_{\theta'},E_{z'}\right)^{\rm T}$, ${\bf H} = \left(H_{r'},H_{\theta'},H_{z'}\right)^{\rm T}$.
Equation (\ref{maxtel}) can be rewritten as:
\begin{align}
  \dfrac{1}{r'} \dfrac{\partial}{\partial r'} (r' \ep_{\theta'}^{(1)} \dfrac{\partial E_{z'}}{\partial r'})
  +\dfrac{1}{r'^2} \dfrac{\partial}{\partial \theta'} (\ep_{r'}^{(1)} \dfrac{\partial E_{z'}}{\partial \theta'})
  &-\dfrac{i}{r'} \dfrac{\partial}{\partial r'} (r'\xi_{\theta'}^{(1)} \dfrac{\partial H_{z'}}{\partial r'})
  -\dfrac{i}{r'^2} \dfrac{\partial}{\partial \theta'} (\xi_{r'}^{(1)} \dfrac{\partial H_{z'}}{\partial \theta'}) \notag \\
  &=-\omega^2 \ep_{z'} E_{z'} -i \omega^2 \xi_{z'} H_{z'}  \notag  \\
  \dfrac{i}{r'} \dfrac{\partial}{\partial r'} (r' \xi_{\theta'}^{(1)} \dfrac{\partial E_{z'}}{\partial r'})
  +\dfrac{i}{r'^2} \dfrac{\partial}{\partial \theta'} (\xi_{r'}^{(1)} \dfrac{\partial E_{z'}}{\partial \theta'})
  &+\dfrac{1}{r'} \dfrac{\partial}{\partial r'} (r'\mu_{\theta'}^{(1)} \dfrac{\partial H_{z'}}{\partial r'})
  +\dfrac{1}{r'^2} \dfrac{\partial}{\partial \theta'} (\mu_{r'}^{(1)} \dfrac{\partial H_{z'}}{\partial \theta'}) \notag \\
  &=i \omega^2 \xi_{z'} E_{z'} - \omega^2 \mu_{z'} H_{z'}
\label{maxcyc}
\end{align}
with $v_{r'}^{(1)} =v_{r'}/{(\ep_{r'} \mu_{r'}-\xi_{r'}^2)}$, $v_{\theta'}^{(1)} =v_{\theta'}/{(\ep_{\theta'} \mu_{\theta'}-\xi_{\theta'}^2)}$.

According to (\ref{cartpol}), the permittivity, permeability and magneto-electric coupling tensors can be expressed in the polar eigenbasis of the metric tensor as
\begin{equation}
 {\bf diag}\,(v_{r'},v_{\theta'},v_{z'})=v_0 {\bf diag}\,(\dfrac{g}{g'r'},\dfrac{g'r'}{g},\dfrac{g'g}{r'})
\label{para2}
\end{equation}
Let us now substitute this formula into (\ref{maxcyc}), we obtain:
\begin{equation}
\begin{array}{l}
  \ep_0\left[\dfrac{\partial}{\partial r'} (\dfrac{g}{g'}\dfrac{\partial E_{z'}}{\partial r'})+\dfrac{g'}{g}\dfrac{\partial^2 E_{z'}}{\partial {\theta'}^2}+\omega^2 a gg' E_{z'}\right]
  -i\xi_0\left[\dfrac{\partial}{\partial r'} (\dfrac{g}{g'} \dfrac{\partial H_{z'}}{\partial r'})+\dfrac{g'}{g}\dfrac{\partial^2 H_{z'}}{\partial {\theta'}^2}-\omega^2 a gg' H_{z'}\right]=0 \\[2.5mm]
  i\xi_0\left[\dfrac{\partial}{\partial r'} (\dfrac{g}{g'}\dfrac{\partial E_{z'}}{\partial r'})+\dfrac{g'}{g}\dfrac{\partial^2 E_{z'}}{\partial {\theta'}^2}-\omega^2 a gg' E_{z'}\right]
  +\mu_0\left[\dfrac{\partial}{\partial r'} (\dfrac{g}{g'}\dfrac{\partial H_{z'}}{\partial r'})+\dfrac{g'}{g}\dfrac{\partial^2 H_{z'}}{\partial {\theta'}^2}+\omega^2 a gg' H_{z'}\right]=0
\end{array}
\label{ehcy1}
\end{equation}
with $a=\ep_0\mu_0-\xi_0^2$. The wave solution for (\ref{ehcy1}) can be expressed as a combination of cylindrical Bessel and Hankel functions $J_m$ and $H^{(1)}_m$ \cite{Kluskens91}
\begin{equation}
\begin{array}{ll}
   r> r_s:& E_{z'}(r',\theta')=\sum\limits_{m}\left[J_m (k_{\pm}g(r'))+a_{m\pm} H^{(1)}_m (k_{\pm}g(r'))\right]e^{im\theta'}  \\
   r_c < r \leq r_s: &E_{z'}(r',\theta')= \sum\limits_{m}\left[b_{m\pm} J_m(k_{\pm}g(r'))+c_{m\pm} H^{(1)}_m(k_{\pm}g(r'))\right]e^{im\theta'} \\
   r < r_s: &E_{z'}(r',\theta')=\sum\limits_{m} d_{m\pm} J_m(k_{\pm}g(r'))e^{im\theta'}
\end{array}
\label{JK}
\end{equation}
where an incident electric field parallel to the cylindrical axis ($z'$-axis) is assumed in the matrix, and $k_{\pm}=\omega(\sqrt{\ep_0\mu_0}\pm \xi_0)$ is the wave number where the subscript $\pm$ stands for the right and left polarized wave in bianisotropic medium. The coefficients $a_{m\pm}$, $b_{m\pm}$, $c_{m\pm}$ and $d_{m\pm}$ can be fixed according to the boundary conditions: Tangential electric and magnetic fields should be continuous across each interface.
\begin{figure}[!htb]
    \centering{
    \subfigure[]{
    \includegraphics[scale=0.45]{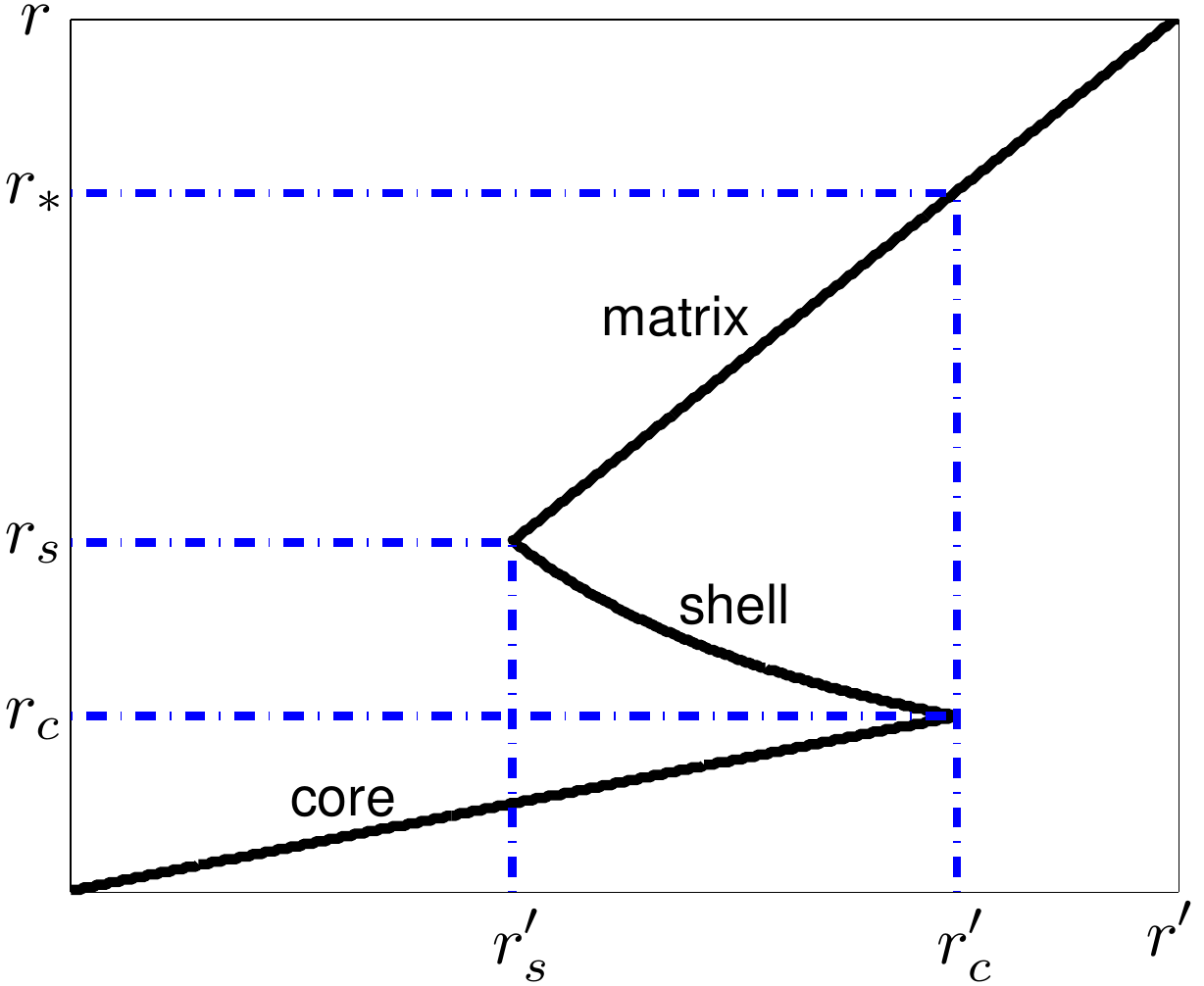}}
    \subfigure[]{
    \includegraphics[scale=0.45]{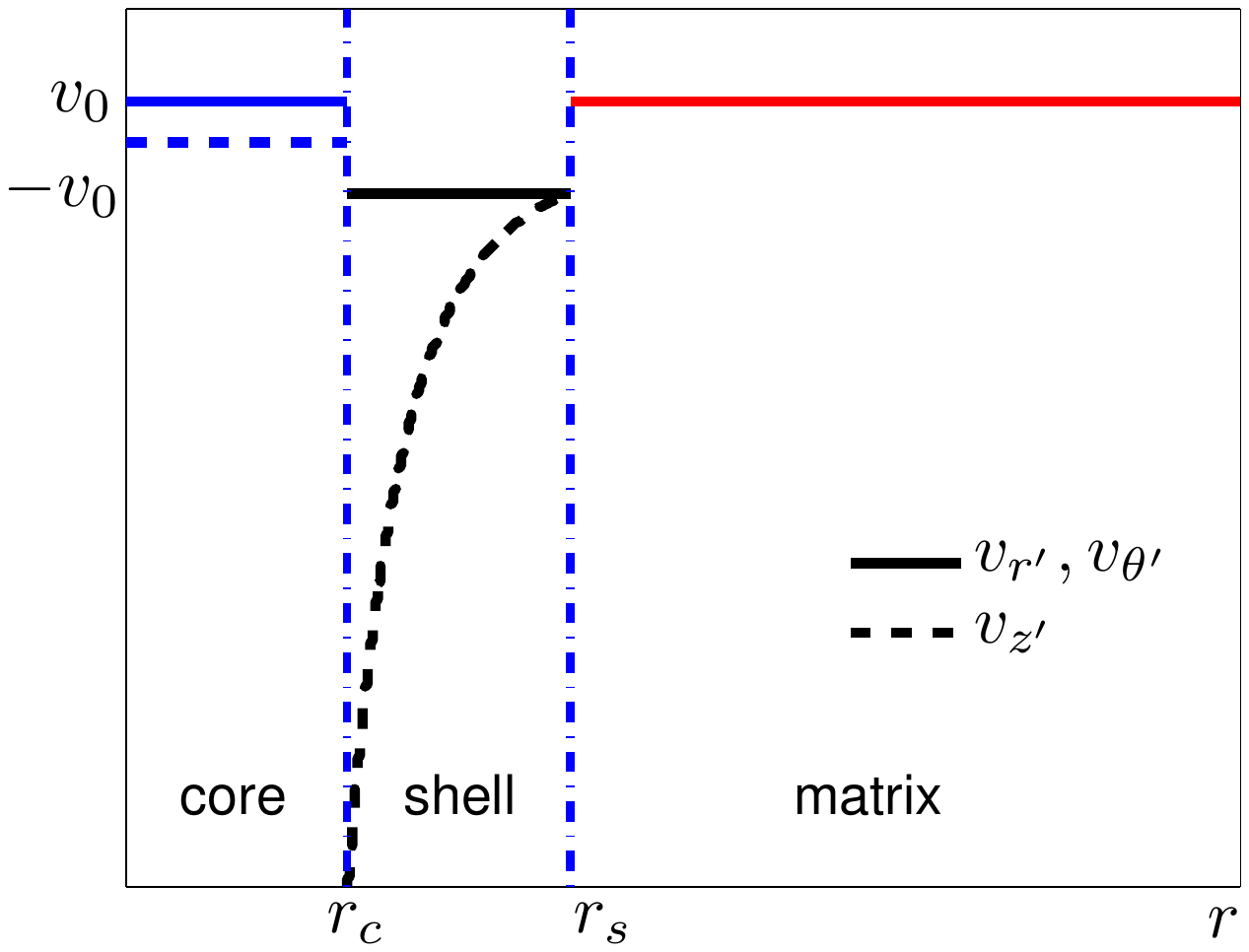}}}
    \caption{(a) Graph of $g(r')$ versus radius $r'$, where parameters in vertical are $r_c=0.02 {\rm m}$ (core radius), $r_s=0.04 {\rm m}$ (shell radius), and $r_*=r_s^2/r_c=0.08 {\rm m}$, $r'_c=r_*$.(b) Curves of parameters defined in (\ref{para2}) versus $r$ in the three regions.}
    \label{figure3}
\end{figure}

Fig. \ref{figure3}(a) shows the mapping from unfolded system $(r,\theta,z)$ to folded system $(r',\theta',z')$, wherein the space overlaps itself but without intersection \cite{Milton08}: Starting from the origin, one first moves in the core with increasing radius until one encounters the core radius $r'_c$, then one moves into the shell with decreasing radius until one reaches the shell radius $r'_s<r'_c$, at which point one moves into the matrix with increasing radius again. By choosing a proper function $g$, folded space can be achieved through a negative slope in region $r'_s<r'<r'_c$, hence a negatively refracting index medium in the shell: The electromagnetic field inside the shell (folded region) is then equal to that in the matrix in this region. Continuity of the radial mapping function is required in order to achieve impedance-matched material interfaces.

To design a SS as in (\ref{para}), we take
\begin{equation}
  r=g(r')=\left\{
  \begin{array}{ll}
  r' r^2_c/r_s^2, & r \leq r_c \\[1mm]
  r_s^2/r', & r_c < r \leq r_s \\
  r', & r >r_s
  \end{array}
  \right.
  \label{f}
\end{equation}
which leads to parameters in (\ref{para2}) as in Fig. \ref{figure3}(b).

\section{External cloaking at all frequency}
Milton and Nicorovici pointed out in 2006 \cite{Milton06} that anomalous resonance occurring near a lens consisting of complementary media will lead to cloaking effect: When there is a finite collection of polarizable line dipoles near the lens, the resonant field generated by the dipoles acts back onto the dipoles and cancels the field acting on them from outside sources. In actuality, the first cloaking device of this type (now known as external cloaking) was introduced by the former authors with McPhedran back in 1994 \cite{Nicorovici94}, but for a different purpose. The first numerical study of this quasi-static cloaking appeared in \cite{Milton08}, and it was subsequently realized that this is associated with some optical space folding \cite{Milton05}.

More precisely, for a Veselago lens \cite{Veselago68} consisting of a slab of thickness $d$ with $\ep=-1$ and $\mu=-1$, surrounded by a medium with $\ep=1$ and $\mu=1$, Milton et al \cite{Milton05} showed that: for a polarizable dipolar line source or single constant energy line source locating at a distance $d_0$ in front of the slab, when $d_0 < d/2$, then the source will be cloaked. However, for a cylindrical lens, the proof of the cloaking phenomenon in the quasi-static limit relies upon the fact that when the permittivities in the core ($\ep_c$), shell ($\ep_s$) and surrounding matrix ($\ep_m$) satisfy $\ep_s\approx-\ep_m\approx-\ep_c$, the collection of polarizable line dipoles lying within a specific distance from the cylindrical lens will be cloaked, with various examples being analyzed in \cite{Milton06}.

\begin{figure}[!htb]
\centering
    \includegraphics[width=0.7\textwidth]{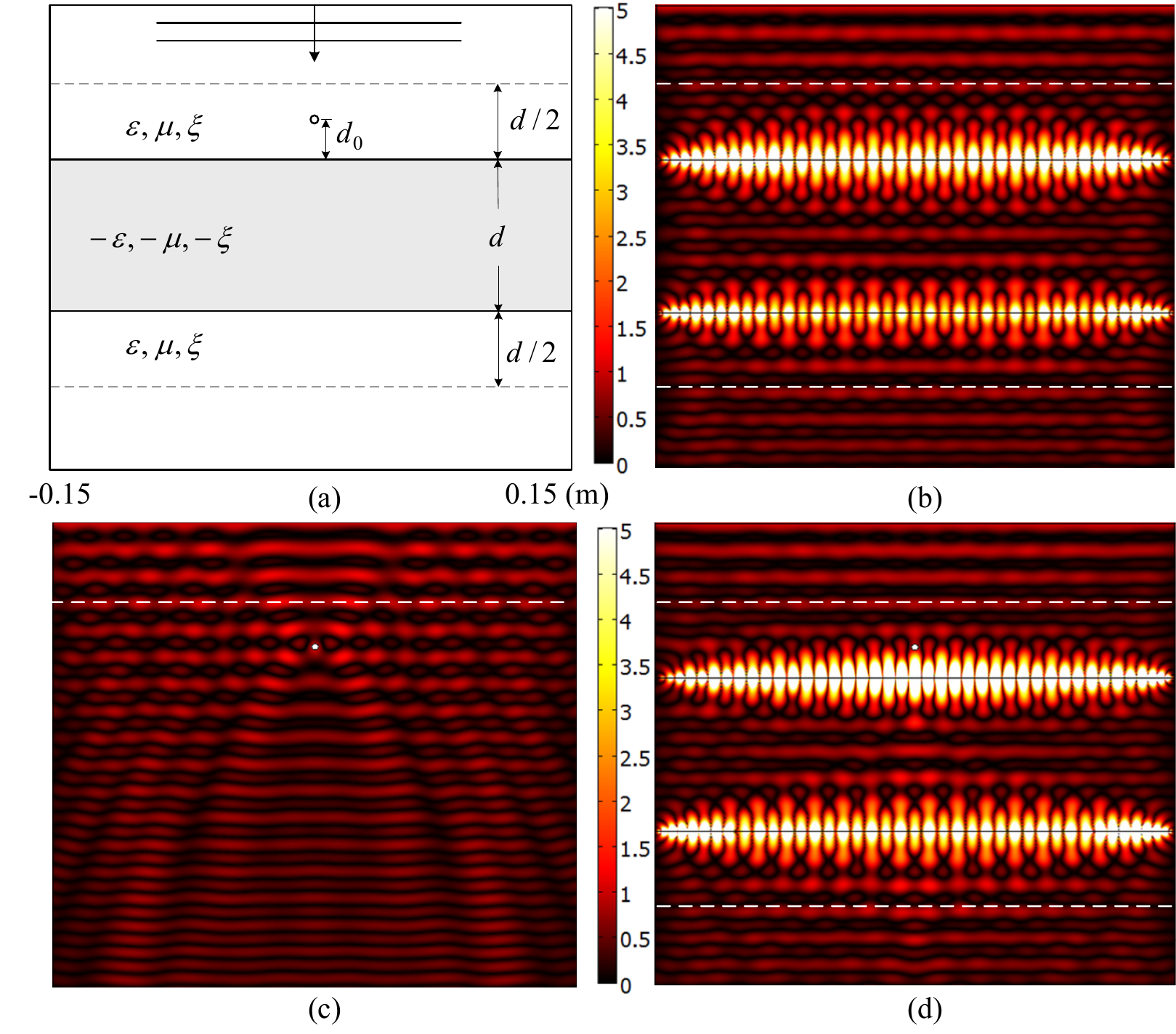}
    \caption{(a) Schematic diagram of a slab lens ($d=0.1$m), and a polarizable line dipole (radius of
    $0.002$m), which is located at a distance $d_0$ above the slab lens in the cloaking region highlighted by dashed line. The parameters in the upper and lower regions are $\ep=\ep_0 I$, $\mu=\mu_0 I$ and $\xi=0.99/c_0 I$, while for the slab they are $\ep=(-\ep_0+i\delta) I$, $\mu=-\mu_0 I$ and $\xi=-0.99/c_0 I$, $I$ is the $3\times 3$ identity matrix; Plots of $\sqrt{E_z^2+H_z^2}$ for: (b) A transparent slab lens; (c) When a dipole line source lies in a background with $\ep=\ep_0 I$, $\mu=\mu_0 I$ and $\xi=0.99/c_0 I$; (d) A dipole located at a distance $d_0=0.02$m (in the cloaking region) from the slab. A TE polarized plane wave is assumed to be incident from above, with a frequency of $8.7$GHz, and $\delta=10^{-15}$. }
    \label{figlenscloak}
\end{figure}

Similar ideas are investigated here for both the slab lens and cylindrical lens, which are all made of bianisotropic media. However, we stress that in our case we would like to achieve external cloaking at finite frequencies. For this, we first consider a slab lens as shown in Fig. \ref{figlenscloak}(a), the thickness of the slab is $d=0.1$m. The parameters in the slab are $\ep=(-\ep_0+i\delta)I$, $\mu=-\mu_0 I$ and $\xi=\xi=-0.99/c_0 I$, while the upper and lower regions are the complementary bianisotropic medium with positive parameters. Assuming a TE polarized plane wave from above, the frequency is $8.7$GHz, the plot of $\sqrt{E_z^2+H_z^2}$ for the slab lens is depicted in Fig. \ref{figlenscloak}(b), a small absorption $\delta=10^{-15}$ has been introduced as the imaginary part of the permittivity of the slab to improve the convergence of the simulation. Fig. \ref{figlenscloak}(c) shows the plot of $\sqrt{E_z^2+H_z^2}$ when a single line dipole (the radius is $0.002$m) locates in the bianisotropic background, a significant perturbation of the field can be observed. However, if we place such kind of dipole at a distance $d_0=0.02$m above the slab, we can see that it is indeed cloaked as shown in Fig. \ref{figlenscloak}(d). As discussed in \cite{yanl13}, the extension of Pendry-Ramakrishna generalized lens theorem \cite{Pendry03} to bianisotropic media shows that: A pair of complementary media makes a vanishing optical path, i.e. the region between the two dashed lines in Fig. \ref{figlenscloak}(a) behaves as though it had zero thickness. This is checked again by comparing the distribution of the fields along the dashed lines in the regions above and below of panels (b) and (d) in Fig. \ref{figlenscloak}.

\begin{figure}[!htb]
    \centering
    \includegraphics[width=0.7\textwidth]{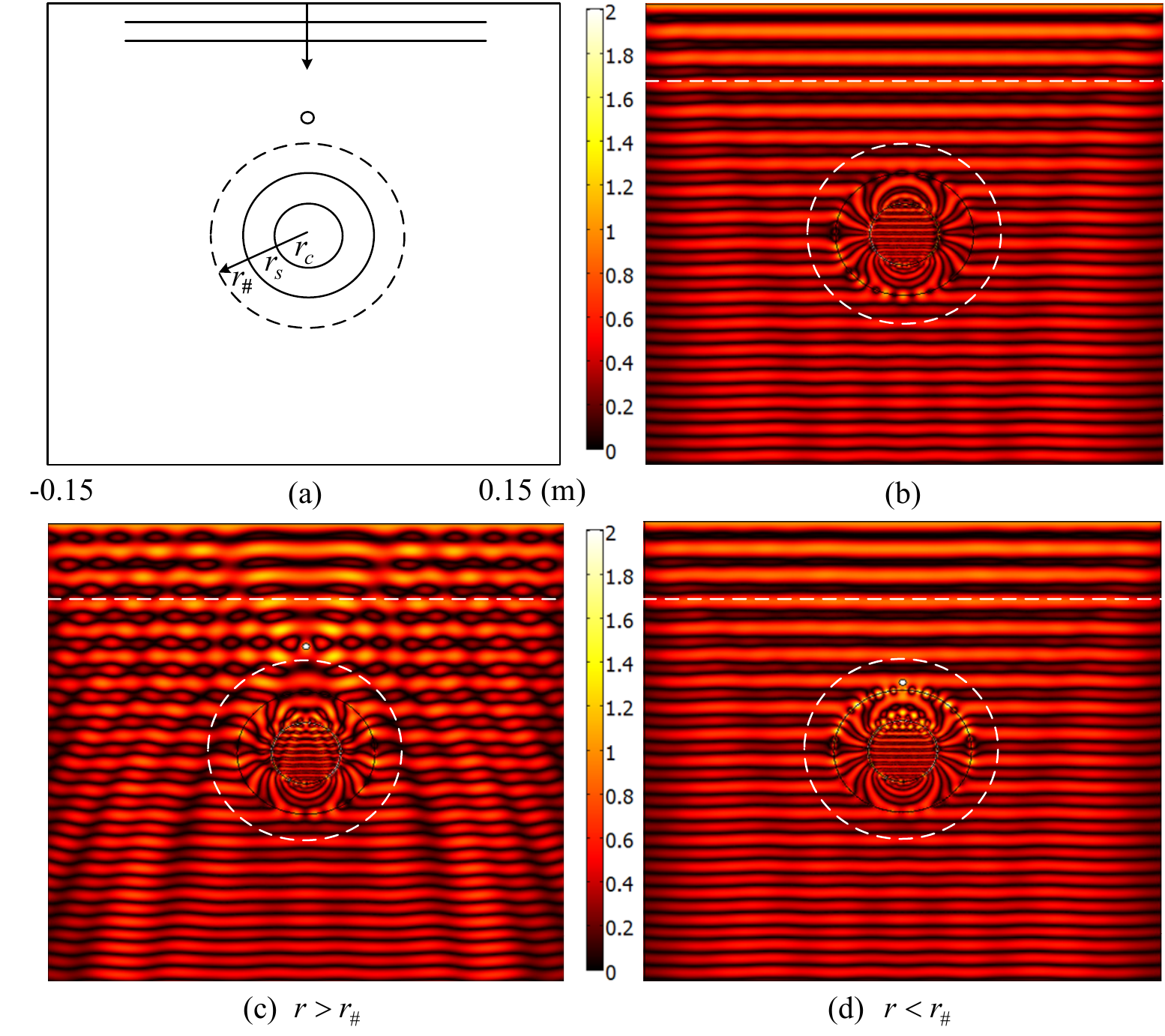}
    \caption{(a) Diagram of a bianisotropic cylindrical lens ($r_c=0.02$m, $r_s=0.04$m) and a polarizable line dipole (radius of $0.002$m). Plots of $\sqrt{E_z^2+H_z^2}$ for: (b) A transparent cylindrical lens without the polarizable dipole; (c) A dipole located outside the cloaking region of radius $r=0.07$m, a significant perturbation of the field can be observed; (d) A dipole located inside the cloaking region of radius $r=0.045$m becomes virtually invisible to the incident field. A TE polarized plane wave incidence with frequency $8.7$GHz is assumed from above.}
    \label{figsd}
\end{figure}

Furthermore, we consider a system consisting of a bianisotropic cylindrical lens and a polarizable line dipole (radius $0.002$m) lying at a distance $r$ from the center point. Parameters of each region are defined by (\ref{para2}) along with (\ref{f}), while a small absorption $\delta=10^{-14}$ is introduced as the imaginary part of permittivity of the shell to ensure numerical convergence of the finite element algorithm. The radii of the cylindrical lens are $r_c=0.02$m, $r_s=0.04$m, respectively. Assuming a TE polarized plane wave with frequency $8.7$GHz from above, the plot of $\sqrt{E_z^2+H_z^2}$ for the cylindrical lens is shown in Fig. \ref{figsd}(b), while Fig. \ref{figsd}(c)-(d) describe the phenomena when the dipole moves towards the cylindrical lens from $r> r_\sharp$ to $r<r_\sharp$ with the cloaking radius $r_\sharp=\sqrt{r_s^3/r_c}=0.0566$m: Cloaking can be observed.

\begin{figure}[!htb]
    \centering
    \includegraphics[width=0.7\textwidth]{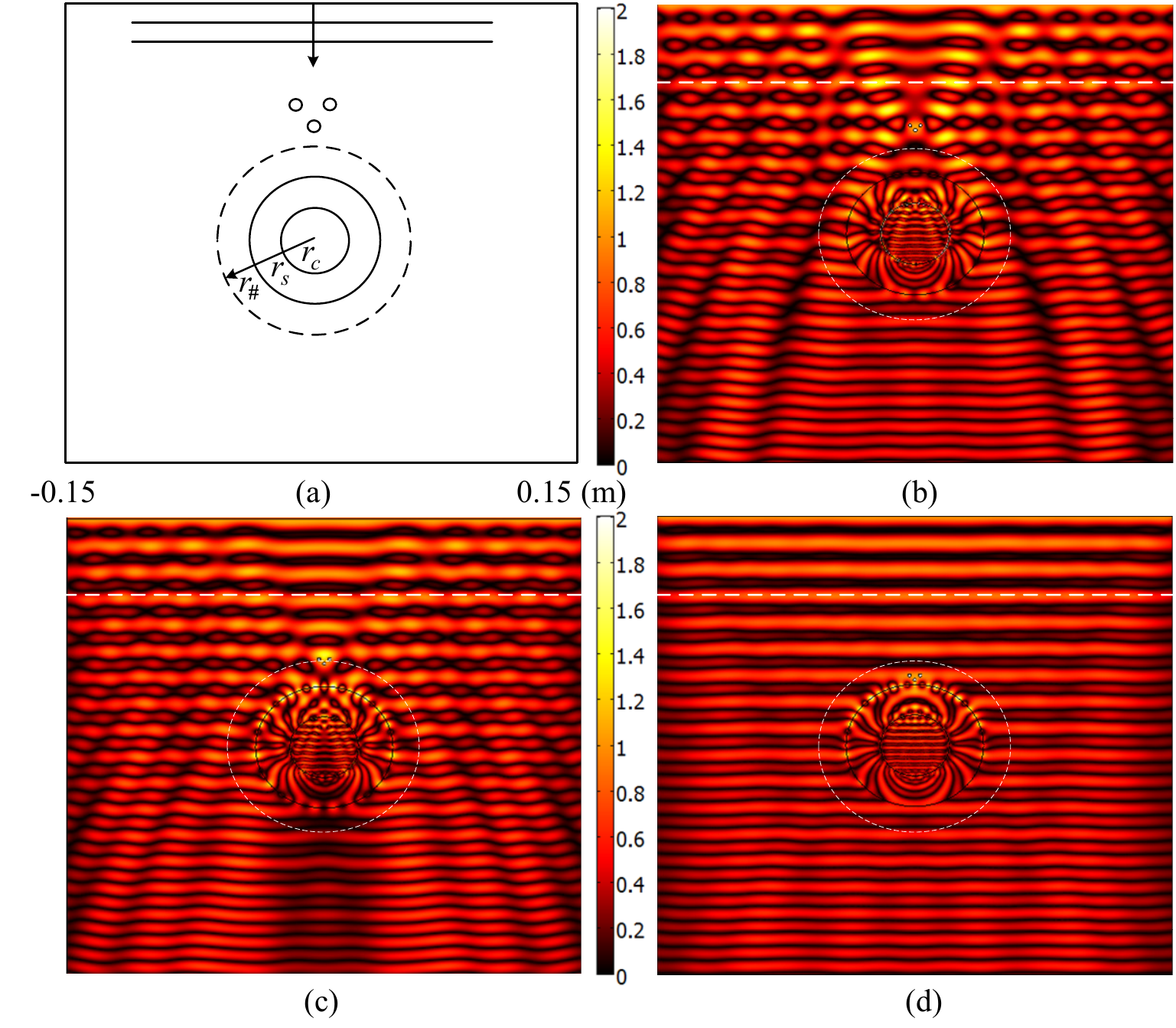}
    \caption{(a) Diagram of a bianisotropic cylindrical lens as in Fig. \ref{figsd} and a triangular polarizable set of line dipoles. Plots of $\sqrt{E_z^2+H_z^2}$ for: (b) Triangular dipoles are totally outside the cloaking region of radius $r_\sharp=\sqrt{r_s^3/r_c}$; (c) The upper two dipoles of triangle are outside while the lower one is inside the cloaking region; (d) All dipoles sit inside the disc of radius $r_\sharp$. A TE polarized plane wave incidence with frequency $8.7$GHz is assumed from above, and $r_s$, $r_c$ and $r_\sharp$ have same values as in Fig. \ref{figsd}. }
    \label{figsvd}
\end{figure}

Moreover, an interplay of a triangular polarizable set of line dipoles with the cylindrical lens is shown in Fig. \ref{figsvd}, the radius of each dipole is $0.001$m, the center-to-center spacing between the upper two dipoles is $0.006$m and the curves connecting their centers form an isosceles right triangle. Fig. \ref{figsvd}(b) shows what happens when all three dipoles are outside the cloaking region; when the lower dipole in the triangle is moving into the cloaking region $r<r_\sharp$, while the upper two are outside, the distribution of the EM field is depicted in Fig. \ref{figsvd}(c); external cloaking is more pronounced when all dipoles enter the cloaking region, see Fig. \ref{figsvd}(d). However, cloaking deteriorates with an increasing number of dipoles, which suggests it would not hold for finite bodies \cite{Bruno07}, which is reminiscent of the ostrich effect \cite{Nicorovici08}. We will come back to the ostrich effect in the next section.

\begin{figure}[!htb]
    \centering
    \subfigure[]{
    \includegraphics[width=0.45\textwidth]{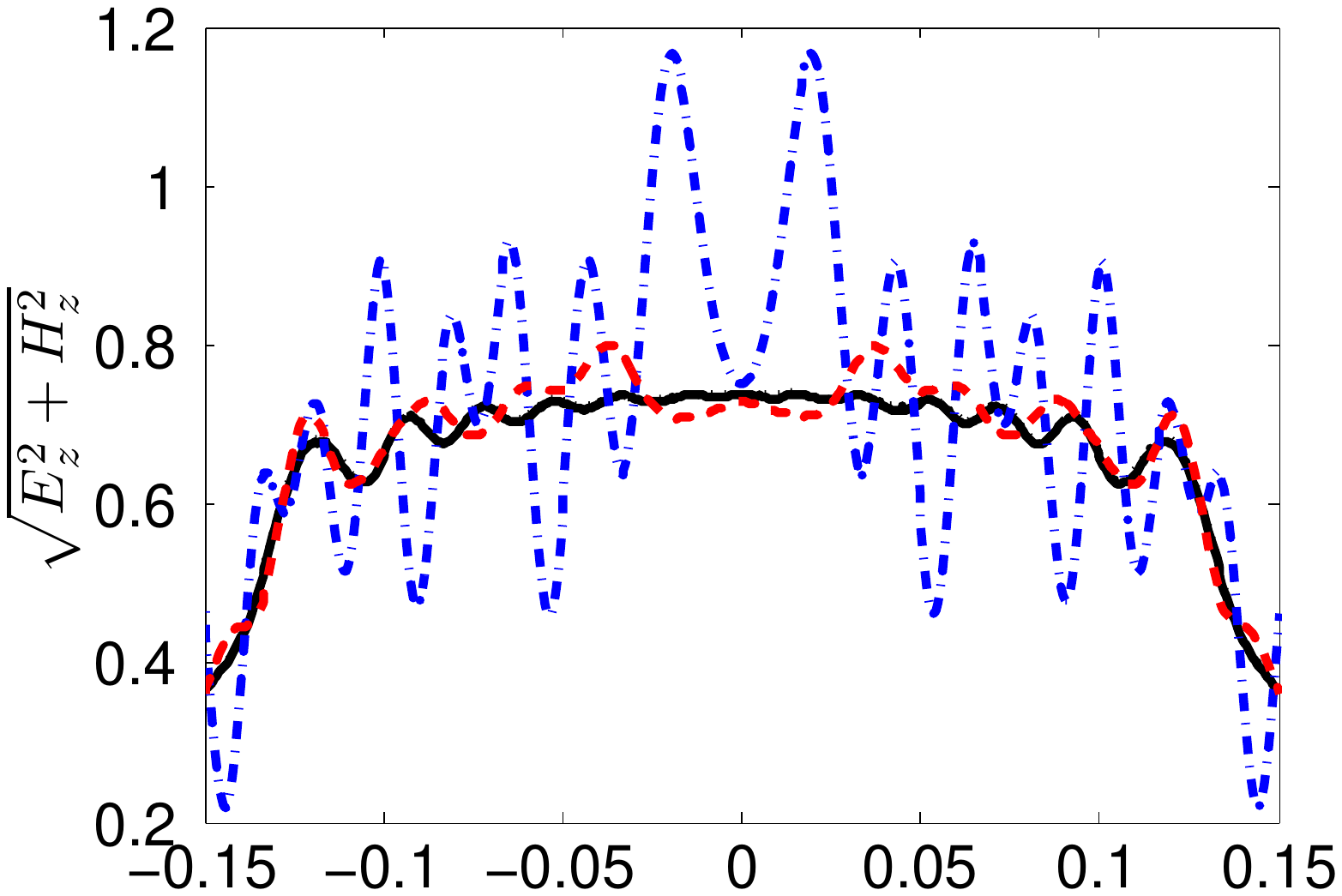}}
    \subfigure[]{
    \includegraphics[width=0.45\textwidth]{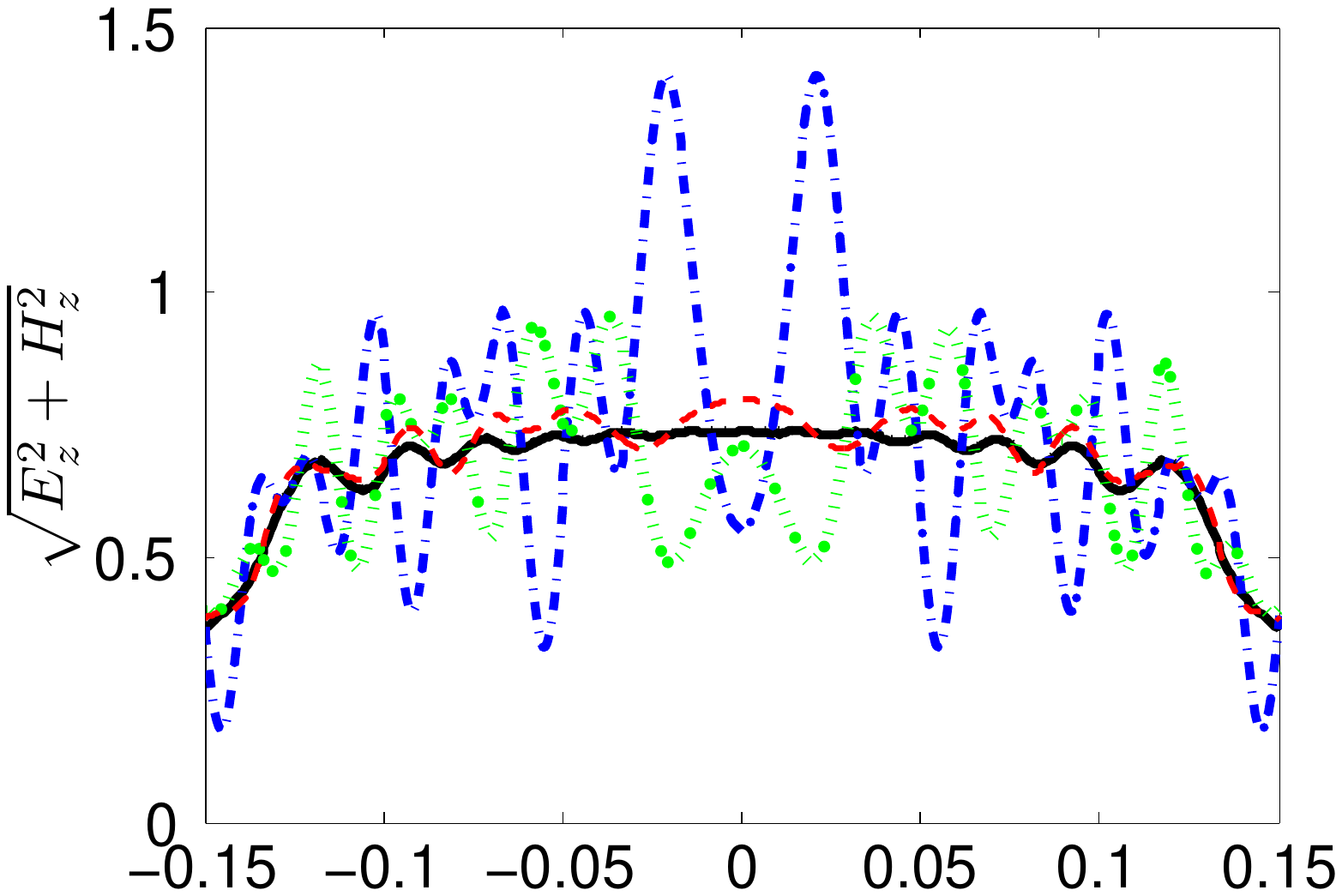}}
    \caption{(a) Comparison of the plots of $\sqrt{E_z^2+H_z^2}$ on the intercepting line for Fig. \ref{figsd}(b)-(d) in black curve, blue dotted-dashed curve and red dashed curve, respectively. (b) Comparison of the plots of $\sqrt{E_z^2+H_z^2}$ on the intercepting line for Fig. \ref{figsvd}(b)-(d) in blue dotted-dashed curve, green dotted and red dashed curve, respectively, the black curve corresponds to the distribution of EM field along the intercepting line for a transparent cylindrical lens without polarizable dipoles.}
    \label{figcutsd}
\end{figure}

As a quantitative illustration, Fig. \ref{figcutsd}(a) shows the distribution of EM field along the intercepting lines in Fig. \ref{figsd}(b)-(d) by black solid, blue dotted-dashed and red dashed curves, respectively. Comparing with the dotted black curve, which is the distribution of EM field in a full bianisotropic background, the black solid one totally matches it, i.e. the cylindrical lens is transparent with respect to the incident wave. Although the red dashed curve does not coincide with the black solid one, it somehow achieves the cloaking effect when the dipole lies inside the cloaking region, by comparison with the blue dotted-dashed curve when the dipole is outside the cloaking region. Similarly, Fig. \ref{figcutsd}(b) shows the distribution of EM field along the intercepting line in Fig. \ref{figsvd}(b)-(d) by blue dotted-dashed curve, green dotted and red dashed curve, respectively; while the black solid curve for a transparent cylindrical lens without polarizable dipoles is the benchmark. Again, an improved scattering EM can be achieved by moving the dipoles into the cloaking region.

\begin{figure}[!htb]
\centering
    \includegraphics[width=0.8\textwidth]{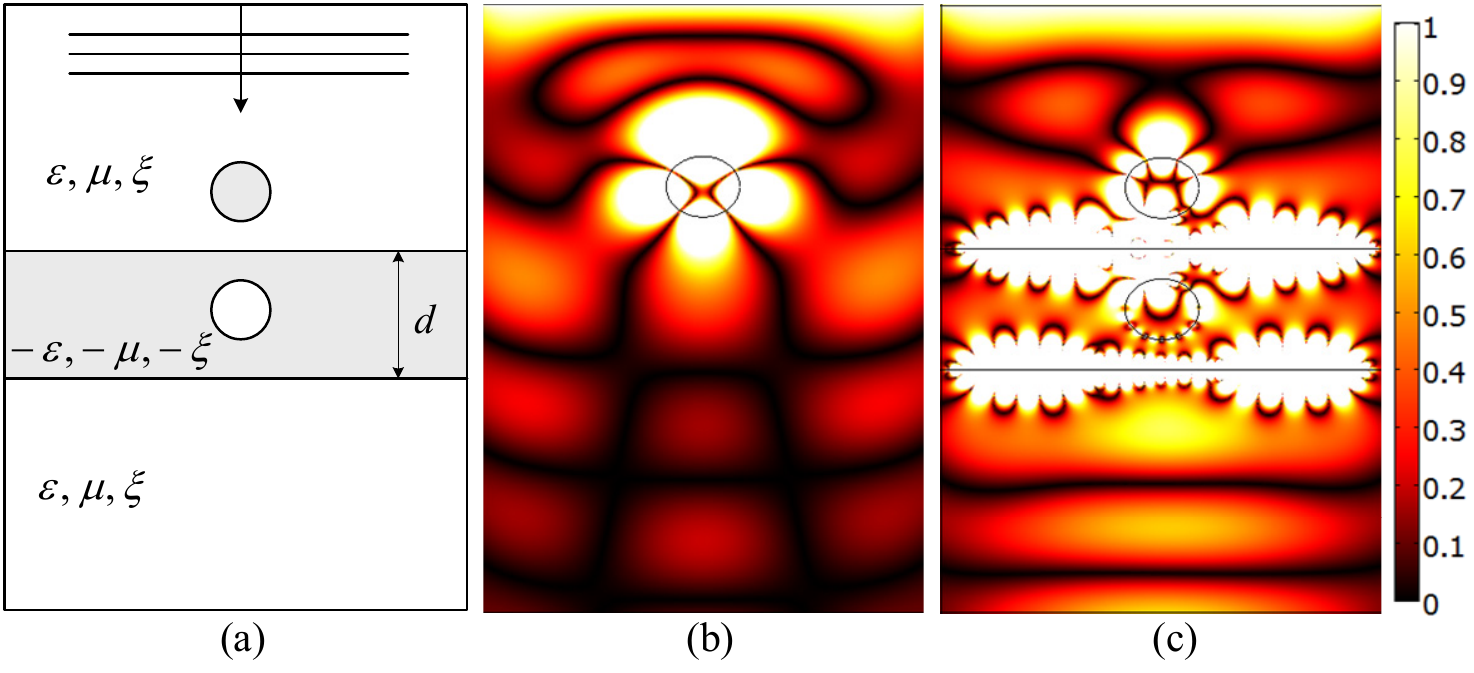}
    \caption{(a) A mirror system made of complementary bianisotropic media, an inclusion with $\varepsilon=\ep_0 I, \mu=\mu_0 I, \xi=0.99/c_0 I$ is placed inside the slab lens with $\varepsilon=(-\ep_0+i\delta) I, \mu=-\mu_0 I, \xi=-0.99/c_0 I$, while a mirror inclusion with $\varepsilon=(-\ep_0+i\delta) I, \mu=-\mu_0 I, \xi=-0.99/c_0 I$ locates in the media with $\varepsilon=\ep_0 I, \mu=\mu_0 I, \xi=0.99/c_0 I$; (b) Plot of $\sqrt{E_z^2+H_z^2}$ for the system wherein only an inclusion with $\varepsilon=(-\ep_0+i\delta) I, \mu=-\mu_0 I, \xi=-0.99/c_0 I$ in the background, a TE polarized plane wave comes from above, the negative inclusion interrupts the propagation of the incident wave; (c) Plot of $\sqrt{E_z^2+H_z^2}$ for the system of (a): waves are transmitted. Absorption is $\delta=10^{-17}$ and the frequency is $1$GHz, the thickness of slab lens is $d=0.1$m, and the radius of the inclusion is $r_0=0.025$m.}
    \label{figlenscut}
\end{figure}

However, this type of external cloaking only works for small polarizable objects (compared to the wavelength) \cite{Bruno07}. If one wishes to cloak a large obstacle, it is possible to resort to an optical paradox put forward by Pendry and Smith \cite{Pendry2003a} in conjunction with the theory of complementary media developed by Pendry and Ramakrishna \cite{Pendry2003}. In the context of complementary bianisotropic media \cite{yanl13}, a circular inclusion consisting of a material with optical parameters $\varepsilon=(-\ep_0+i\delta) I, \mu=-\mu_0 I, \xi=-0.99/c_0 I$ and of radius $r_0$, which is located in a medium with parameters $\varepsilon=\ep_0 I, \mu=\mu_0 I, \xi=0.99/c_0 I$ is optically canceled out by an inclusion with $\varepsilon=\ep_0 I, \mu=\mu_0 I, \xi=0.99/c_0 I$ of same diameter $r_0$ in a slab lens of medium $\varepsilon=(-\ep_0+i\delta) I, \mu=-\mu_0 I, \xi=-0.99/c_0 I$, see Fig. \ref{figlenscut} for $r_0=0.025$m and $\delta=10^{-17}$. The physical interpretation of this striking phenomenon is that some resonances building up in this optical system make possible some tunneling of the EM field through the slab lens and inclusions. One can see on Fig. \ref{figlenscut} that the transmission is nevertheless not perfect, but the forward scattering is much reduced in panel (c), compared to panel (b), and we numerically checked that this kind of optical cancelation breaks down at higher frequencies.
\begin{figure}[!htb]
\centering
    \includegraphics[width=0.85\textwidth]{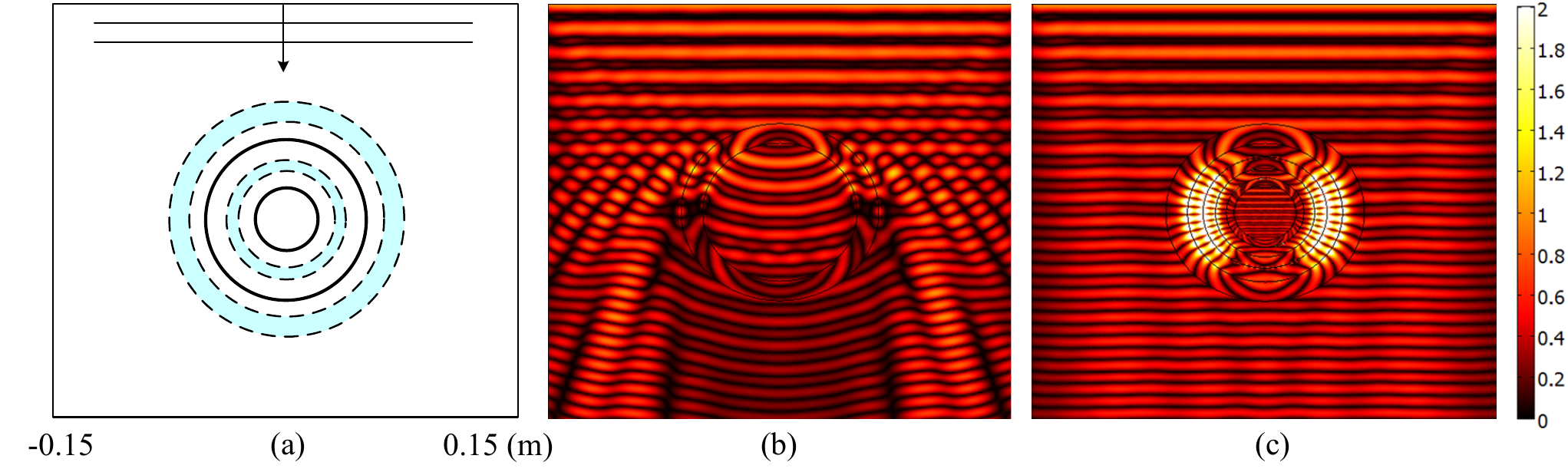}
    \caption{(a) An annulus with $\varepsilon=\ep_0 {\bf diag}(1,1,r^4_s/r^4), \mu=\mu_0 {\bf diag}(1,1,r^4_s/r^4), \xi=0.99/c_0 {\bf diag}(1,1,r^4_s/r^4)$ and of radii $0.025{\rm m}-0.032{\rm m}$ is placed inside the shell of a cylindrical lens as shown in Figs. \ref{figsd}-\ref{figsvd}, while a mirror inclusion with $\varepsilon=(-\ep_0 +i\delta)I, \mu=-\mu_0 I, \xi=-0.99/c_0 I$ and of radii $0.05$m-$0.064$m is placed in the background; (b) Plot of $\sqrt{E_z^2+H_z^2}$ for the optical system with only an inclusion with $\varepsilon=(-\ep_0 +i\delta)I, \mu=-\mu_0 I, \xi=-0.99/c_0 I$ in the background, a TE polarized plane wave comes from above; (c) Plot of $\sqrt{E_z^2+H_z^2}$ for the optical system of (a): waves propagate undisturbed. Absorption is $\delta=10^{-14}$ and the frequency is $8.7$GHz.}
    \label{figcycut}
\end{figure}

Similarly for the cylindrical lens in Figs. \ref{figsd}-\ref{figsvd}, an annulus with parameters $\varepsilon=(-\ep_0+i\delta) I, \mu=-\mu_0 I, \xi=-0.99/c_0 I$ and radii $0.05$m-$0.064$m can be cloaked by an annulus with parameters $\varepsilon=\ep_0 {\bf diag}(1,1,r^4_s/r^4), \mu=\mu_0 {\bf diag}(1,1,r^4_s/r^4), \xi=0.99/c_0 {\bf diag}(1,1,r^4_s/r^4)$ and of radii $0.025$m-$0.032$m, which is located in the shell with $\varepsilon=(-\ep_0+i\delta){\bf diag}(1,1,r^4_s/r^4)$, $\mu=-\mu_0 {\bf diag}(1,1,r^4_s/r^4)$, $\xi=-0.99/c_0 {\bf diag}(1,1,r^4_s/r^4)$, as shown in Fig. \ref{figcycut}(a). Note that the ratios for each component of the parameters of these two complementary annuli are equal to ${\bf diag}(g/g'r',\,g'r'/g,\, g'g/r')$ with coordinate transformation function defined in (\ref{f}); while their radii are also satisfying the relation (\ref{f}). For a TE polarized plane wave with frequency $8.7$GHz coming from above, the distribution of EM field of the optical system with a negative annulus in the background and that of the system in panel (a) are depicted in (b) and (c), respectively. The negative inclusion in the background is cloaked by introducing the negative cylindrical lens with a complementary annulus.

\begin{figure}[!htb]
\centering
    \includegraphics[width=0.85\textwidth]{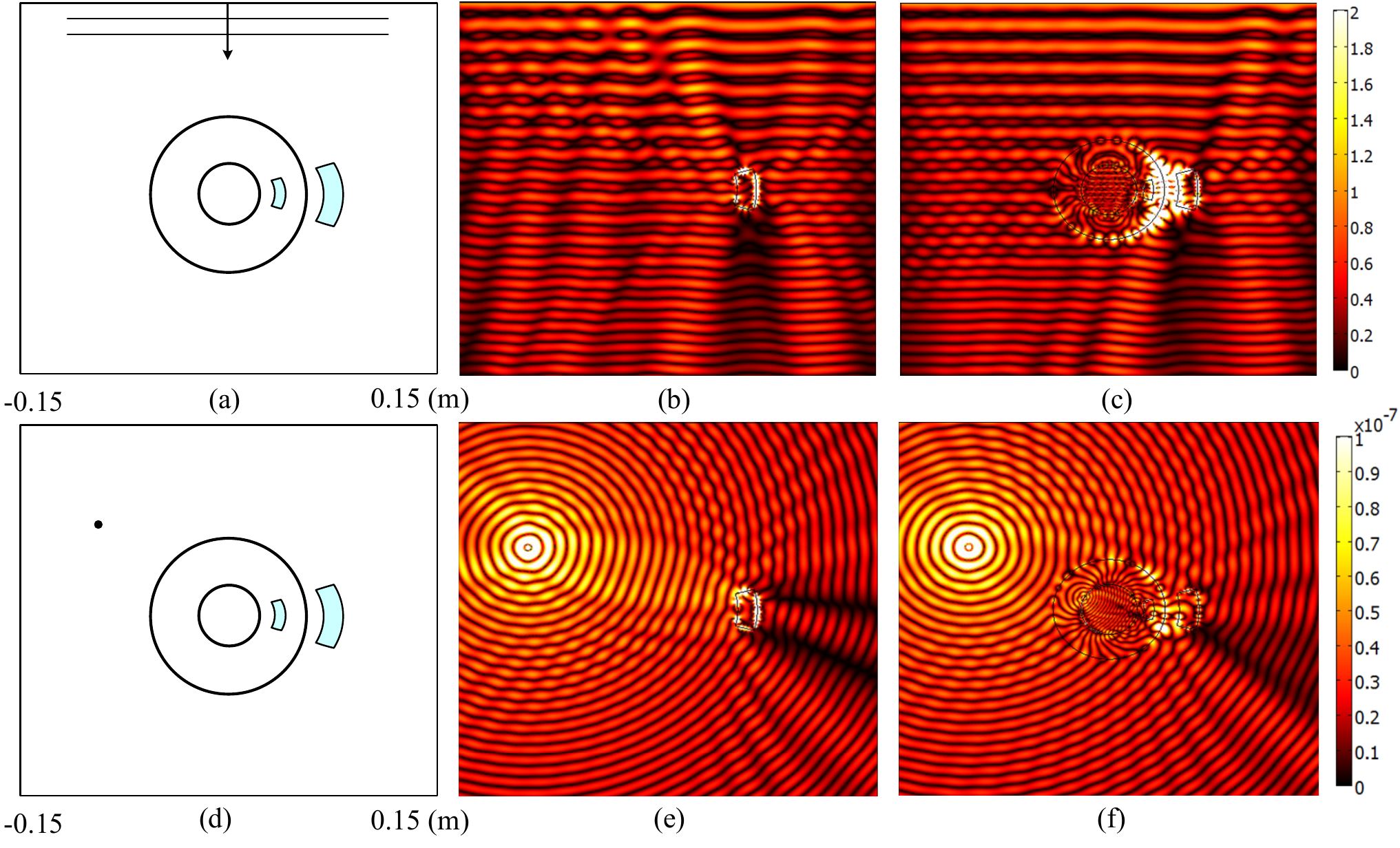}
    \caption{(a) A curved sheet with $\varepsilon=\ep_0 {\bf diag}(1,1,r^4_s/r^4)$, $\mu=\mu_0 {\bf diag}(1,1,r^4_s/r^4)$, $\xi=0.99/c_0 {\bf diag}(1,1,r^4_s/r^4)$ and of radii $0.025$m-$0.032$m is placed inside the shell with $\varepsilon=(-\ep_0+i\delta){\bf diag}(1,1,r^4_s/r^4)$, $\mu=-\mu_0 {\bf diag}(1,1,r^4_s/r^4)$, $\xi=-0.99/c_0 {\bf diag}(1,1,r^4_s/r^4)$, while a mirror sheet with $\varepsilon=(-\ep_0 +i\delta)I, \mu=-\mu_0 I, \xi=-0.99/c_0 I$ and of radii $0.05$m-$0.064$m locates in the background with $\varepsilon=\ep_0 I$, $\mu=\mu_0 I$, $\xi=0.99/c_0 I$; Plots of $\sqrt{E_z^2+H_z^2}$ for: (b) A curved sheet with $\varepsilon=(-\ep_0 +i\delta)I$, $\mu=-\mu_0 I$, $\xi=-0.99/c_0 I$ in the background; (c) The system of (a): propagation of the incident wave is improved. Absorption is $\delta=10^{-14}$ and the frequency is $8.7$GHz.}
    \label{figcycut2}
\end{figure}

We then replace the annuli by curved sheets, which are pieces of the two annuli in Fig. \ref{figcycut}(a), as shown in Fig. \ref{figcycut2}(a). The parameters for the structure are preserved as same, the distributions of EM field are depicted in panels (b)-(c). When there is a negative curved sheet located in the background, a scattering effect can be observed in (b), if we introduce the negative shell with a positive sheet to cancel the negative sheet, the scattering arising from the negative sheet is improved. One can nevertheless observe some side-scattering effects, which could be improved if the wave is incident from the left. This suggested us to introduce a line source as shown in Fig. \ref{figcycut2}(d), the distributions of EM field for a negative curved sheet without and with the complementary structure achieved by a negative shell with positive curved sheet, are shown in panels (e)-(f), the cloaking effect for the sheet is even more apparent, especially in view of the much reduced shadow behind the negative curved sheet. Note also that we numerically checked cloaking worsens with higher-frequencies, and improves in the quasi-static limit. We also looked at sheets of more complex shapes, with similar results.

\section{Ostrich effect at low frequency}
\begin{figure}[!htb]
    \centering
    \includegraphics[width=0.85\textwidth]{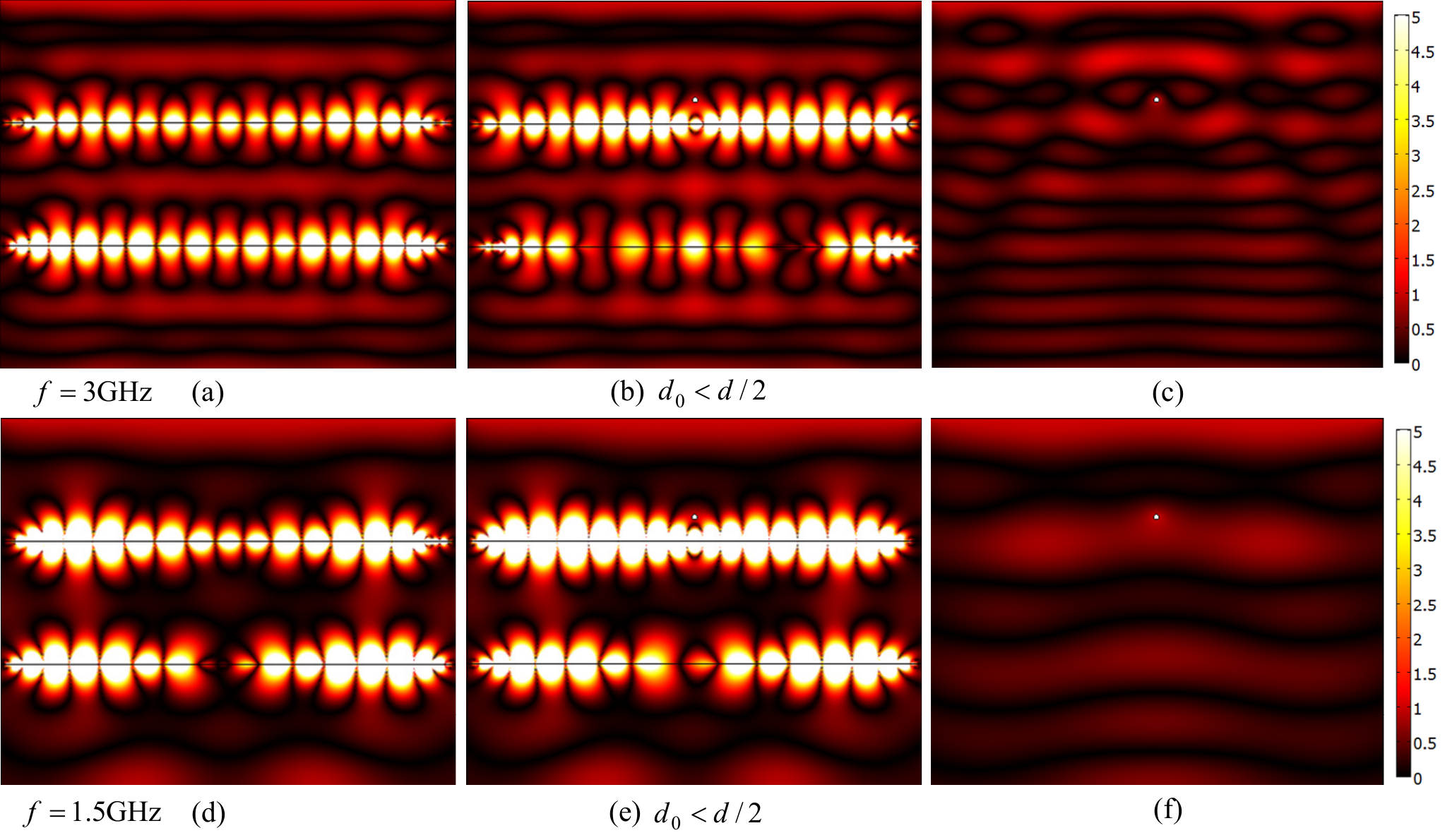}
    \caption{Plots of $\sqrt{E_z^2+H_z^2}$ for the same system as in Fig. \ref{figlenscloak}(a), a TE polarized plane wave incidence with frequency $3$GHz: (a) Scattering property of the bianisotropic slab lens with the thickness $d=0.1$m; (b) Dipole (radius of $0.002$m) is inside the cloaking region with $d_0 < d/2$ ($d_0=0.02$m); (c) The dipole sits inside a bianisotropic background. Similar illustration is indicated in (d)-(f) when the frequency becomes $1.5$GHz.}
    \label{figlo}
\end{figure}

Finally, we numerically checked that if one increases the wavelength of the incident wave to allow the bianisotropic slab lens to become visible, then an ostrich effect can be observed in the system as shown in Fig. \ref{figlenscloak}(a). First, we take the frequency as $f=3$GHz, Fig. \ref{figlo}(a) shows the plot of $\sqrt{E_z^2+H_z^2}$ in a slab lens with thickness $d=0.1$m: The slab lens becoming more visible than Fig. \ref{figlenscloak}(b) by comparing the forward scattering fields in the lower space of the lens; furthermore, if we put a dipole (radius $0.002$m) at a distance $d_0<d/2$ ($d_0=0.02$m), the distribution of the fields is shown in (b), where (c) shows the scattering fields for a dipole in the bianisotropic background. Comparing panels (a) and (b), we can see that the dipole is cloaked, i.e. external cloaking leads to the ostrich effect \cite{Nicorovici08}. Similar effects can be observed by increasing the wavelength of the incidence as shown in Fig. \ref{figlo} (d)-(f) with $f=1.5$GHz.

\begin{figure}[!htb]
    \centering
    \includegraphics[width=0.75\textwidth]{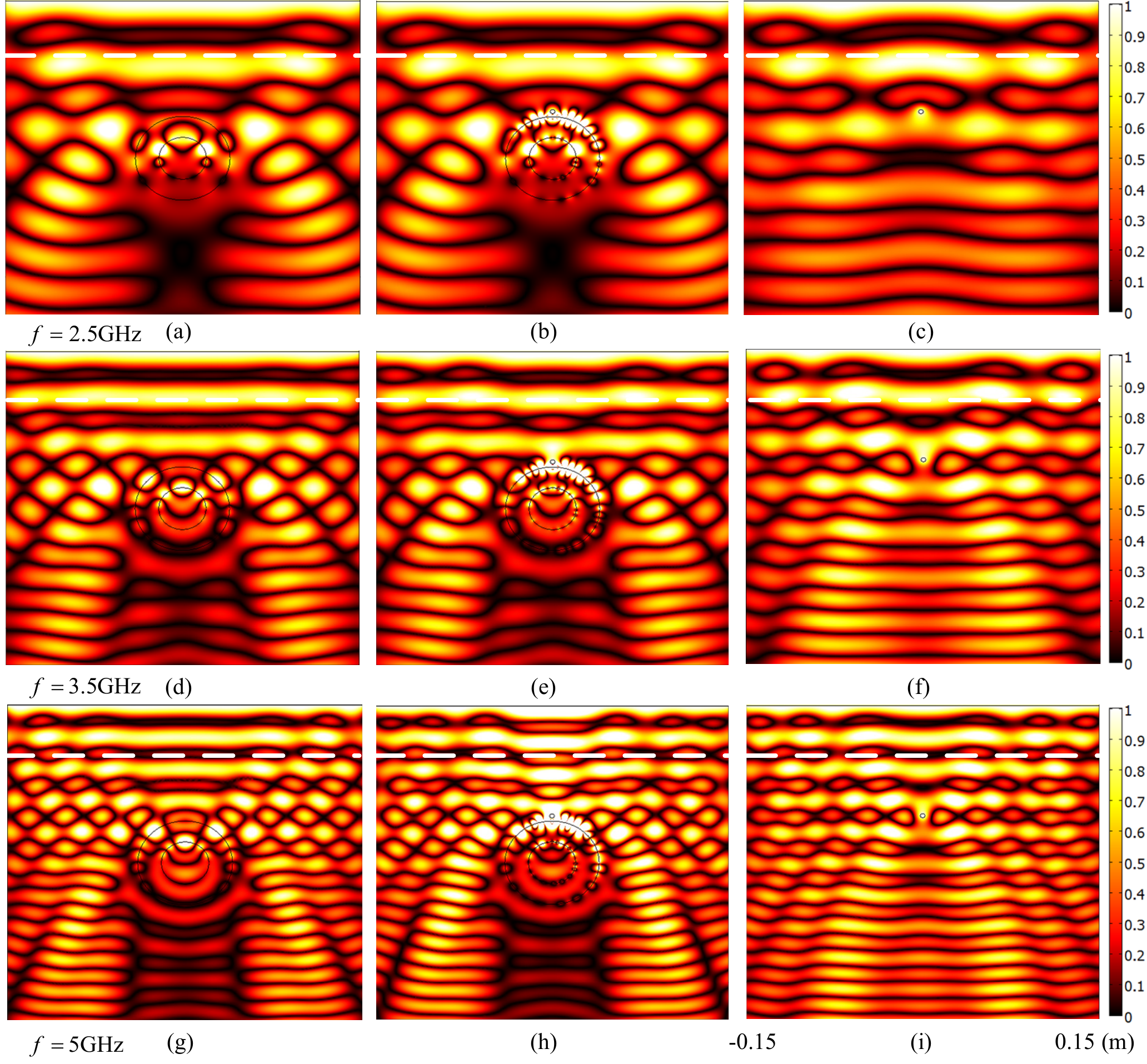}
    \caption{Plots of $\sqrt{E_z^2+H_z^2}$ under a TE polarized incidence with frequency $2.5$GHz: (a) Cylindrical lens with parameters $\ep=(-\ep_0+i\delta)I$, $\mu=-\mu_0I$, $\xi=-0.99/c_0 I$ in the shell, and $\ep=\ep_0I$, $\mu=\mu_0I$, $\xi=0.99/c_0 I$ in both the background and core; a small absorption $\delta=10^{-20}$ is introduced as the imaginary part of permittivity in the shell to improve the convergence of the package COMSOL; (b) A dipole (radius of $0.002$m) locates inside the cloaking region $r<r_\sharp$ ($r=0.045$m); (c) A dipole in the bianisotropic background. The ostrich effect becomes weaker along with an increasing frequency, e.g. panels (d)-(f) for $f=3.5$GHz, and (g)-(i) for $f=5$GHz.}
    \label{figso}
\end{figure}

Meanwhile, we also numerically checked that if one considers a bianisotropic shell with relative permittivity, permeability and magneto-electric coupling tensors all equal to $-I$, and a small absorption $\delta=10^{-20}$ has been introduced as the imaginary part of the permittivity of the shell to improve the convergence of the package COMSOL; while the parameters in the bianisotropic background and core are $v=v_0I$. The frequency of the incidence is assumed to be $f=2.5$GHz allowing a wavelength comparable with the size of the cylindrical lens (same radii as Fig. \ref{figsd}). The numerical illustration for such a lens is shown in Fig. \ref{figso}(a), where the cylindrical lens becomes visible. When we place a dipole (radius of $0.002$m) inside the cloaking region $r<r_\sharp$ ($r=0.045$m), similar distribution of EM field as panel (a) can be observed, i.e. an external cloaking can still be observed, which is the ostrich effect \cite{Nicorovici08}; panel (c) shows the case when there is only a dipole located at the bianisotropic background as a comparison. If one increases the frequency to $f=3.5$GHz even $f=5$GHz, the ostrich effect becomes weak, as shown in panels (d)-(f), (g)-(i).
\begin{figure}[!htb]
    \centering
    \includegraphics[width=0.95\textwidth]{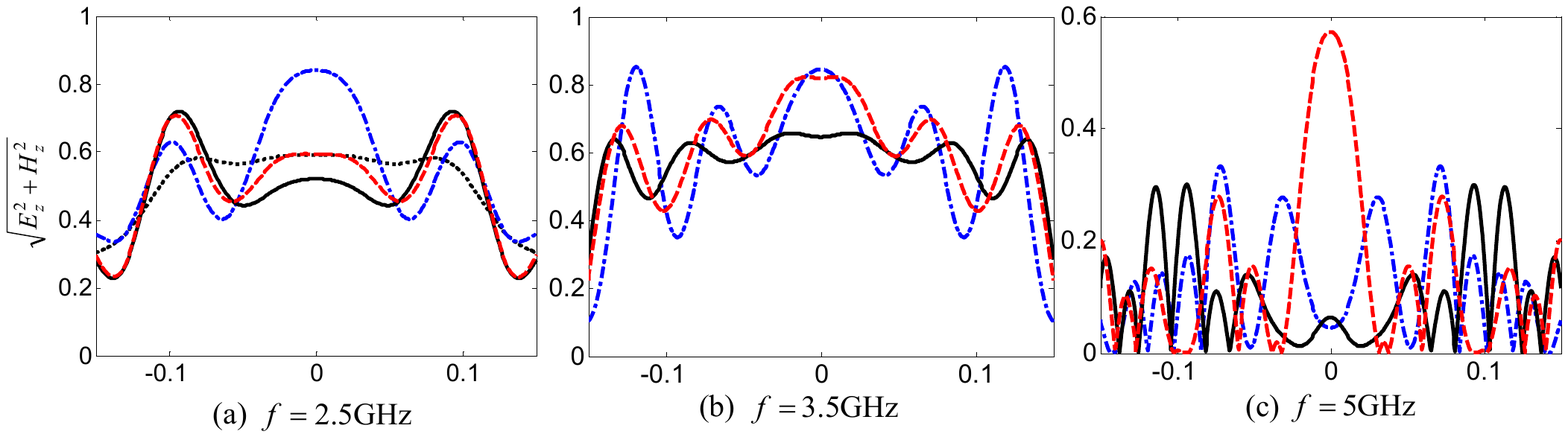}
    \caption{Comparison of the plots of $\sqrt{E_z^2+H_z^2}$ on the intercepting line for Fig. \ref{figso}; Black solid, red dashed and blue dotted-dashed curves are respectively representing the distribution of EM field along the upper intercepting line of the cylindrical lens, lens with a dipole located inside the cloaking region and a single dipole located at the bianisotropic background: (a) $f=2.5$GHz, as a benchmark, the EM field along the intercepting line in bianisotropic background is denoted in dotted black curve; (b) $f=3.5$GHz; (c) $f=5$GHz.}
    \label{figost}
\end{figure}

Again, we compare the distribution of the EM field along the upper intercepting line in the three systems of each line as shown in Fig. \ref{figost}, they are denoted in black solid, red dashed and blue dotted-dashed curves, respectively. At low frequency $f=2.5{\rm GHz}$, the EM field along the intercepting line in bianisotropic background is denoted in dotted black curve as a comparison, the mismatch between it and the black solid curve indicates that the cylindrical lens becomes visible; if we place a dipole in the cloaking region as shown in Fig. \ref{figso}(b), the distribution of EM field of which is marked by dashed curve, a relatively small difference can be observed from the solid line; however, for a dipole located in the background, we have a quite different dotted-dashed line for the distribution of EM field; in other words, an ostrich effect can be achieved. Note that, if we increase the frequency, the phenomenon collapses, see Fig. \ref{figost}(b) and (c) for the frequencies $3.5$GHz and $5$GHz.

\section{External cloaking in air with chiral slab lens}
\begin{figure}[!htb]
    \centering
    \includegraphics[width=0.65\textwidth]{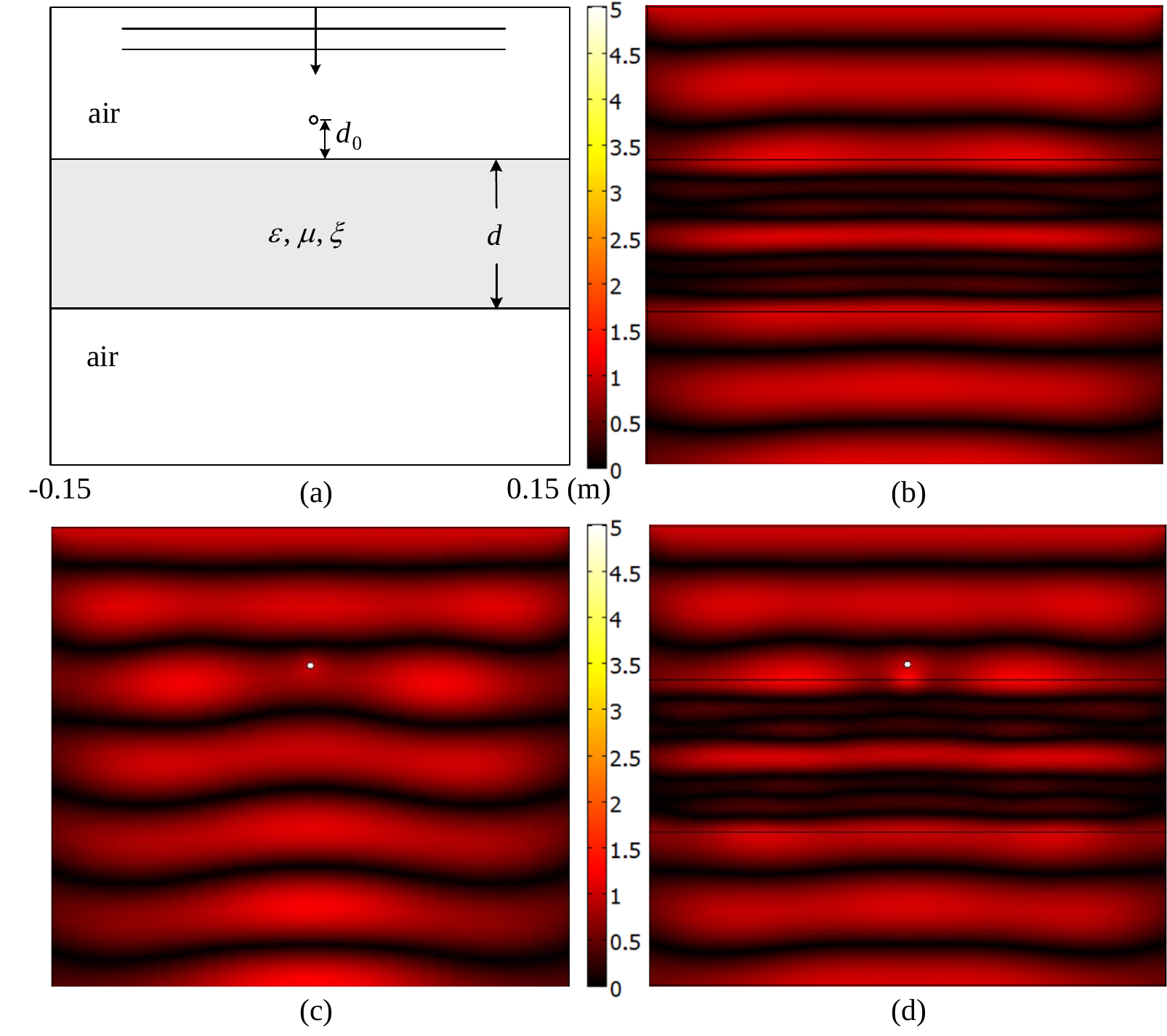}
    \caption{(a) Diagram of a chiral slab lens with $\ep=(\ep_0+i\delta)I$, $\mu=\mu_0I$, $\xi=1.975/c_0I$, and the upper and lower regions are air. Plots of $\sqrt{E_z^2+H_z^2}$ for a chiral slab lens \cite{Jin05}, a TE polarized plane wave incidence with wavelength $\lambda=d=0.1$m, $\delta=10^{-16}$: (b) Scattering property of the chiral slab lens when the wavelength of radiation is comparable with the size of the structure; (c) A dipole is placed in air; when the dipole is placed very close to the chiral slab lens, partial external cloaking (i.e. for the forward scattering) can be observed by comparison of panels (d) and (c).}
    \label{figHe}
\end{figure}

The previous designs involve fairly complex fully bianisotropic media. It is interesting to simplify these parameters in order to foster experimental efforts in this emergent area. We explore the external cloaking effect in air with a chiral slab lens, wherein the slab lens is exactly the same as pointed out by Jin and He \cite{Jin05}, since the chiral slab lens possess a negative refractive index, where the anomalous resonance occurs. Fig. \ref{figHe}(a) shows the chiral slab lens with the upper and lower regions being air, the parameters in the slab are $\ep=(\ep_0+i\delta)I$ with $\delta=10^{-16}$, $\mu=\mu_0I$, $\xi=1.975/c_0I$, and the thickness of the slab is $d=0.1$m. We consider a TE polarized plane wave from above, the wavelength of which is equal to the thickness of the slab as defined in \cite{Jin05}. Panel (b) is the plot of $\sqrt{E_z^2+H_z^2}$ of this chiral slab lens; while (c) shows the distribution of EM field in a system, wherein a dipole with radius $0.002$m located in the air. Since the refractive index of the chiral slab satisfies $n_-\approx-1$, anomalous resonance can occur at the interfaces between the air and the slab, if we put the dipole quite close to the slab at a distance $d_0=0.01$m, then a quasi-cloaking effect can be observed as shown in (d).
\begin{figure}[!htb]
    \centering
    \includegraphics[width=0.8\textwidth]{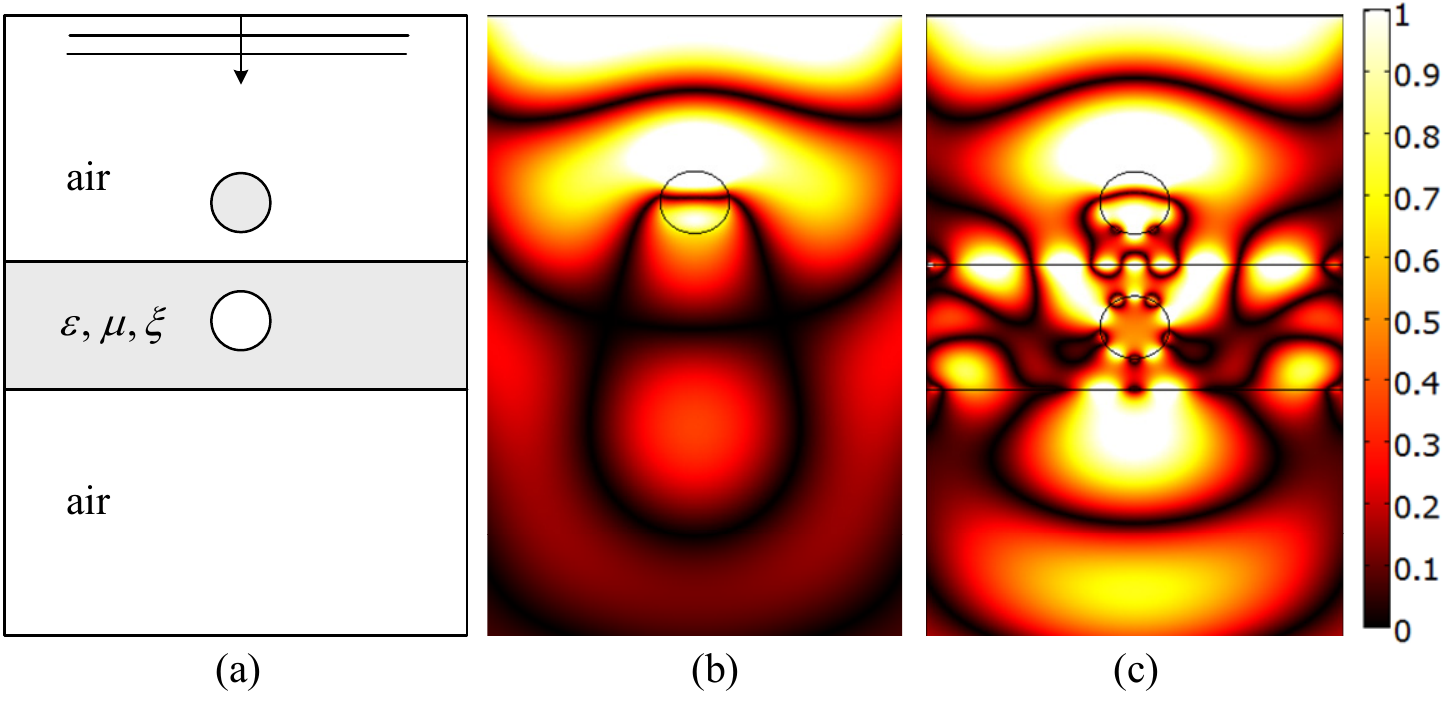}
    \caption{(a) A mirror system made by air and bianisotropic media, a circular inclusion with $\varepsilon=(\ep_0+i\delta) I$, $\mu=\mu_0 I$, $\xi=1.975/c_0 I$ is placed inside the air, while a mirror inclusion of air locates in the media with $\varepsilon=(\ep_0+i\delta)I$, $\mu=\mu_0 I$, $\xi=1.975/c_0 I$; (b) Plot of $\sqrt{E_z^2+H_z^2}$ for the system wherein only an inclusion with $\varepsilon=(\ep_0+i\delta) I$, $\mu=\mu_0 I$, $\xi=1.975/c_0 I$ in air, a TE polarized plane wave comes from above, the inclusion interrupt the propagation of the incidence; (c) Plot of $\sqrt{E_z^2+H_z^2}$ for the system of (a): waves are transmitted. A small absorption $\delta=10^{-16}$ is introduced and the frequency is $1.4$GHz, the thickness of slab lens is $d=0.1$m, and the radius of the inclusion is $0.025$m.}
    \label{figHecut}
\end{figure}

To cloak a large obstacle, we implement the similar idea as Fig. \ref{figlenscut}(a), a circular inclusion with parameters $\varepsilon=(\ep_0+i\delta)I$, $\mu=\mu_0 I$, $\xi=1.975/c_0 I$ and of radius $0.025$m is placed in air, is partly canceled out by an inclusion of air in a chiral slab lens with parameters $\varepsilon=(\ep_0+i\delta)I, \mu=\mu_0 I, \xi=1.975/c_0 I$, since the chiral medium possess a negative index $n_- \approx -1$ which is opposite to the index of air. For a TE polarized incidence, the distribution of the EM field is shown in Fig. \ref{figHecut}, (b) is the case when there is only a chiral inclusion located in air, while (c) shows the reduced scattering (in forward scattering) for the chiral inclusion in air, when a slab is added nearby. The frequency is $1.4$GHz, the thickness of slab lens is $d=0.1$m, and the radius of the inclusion is $0.025$m. A small absorption $\delta=10^{-16}$ is introduced as the imaginary part of the permittivity in chiral medium. Although the transmission of the incident wave is not perfect, it opens us a possible route to the application of the bianisotropic media.  Importantly, bianisotropic media can be achieved from dielectric periodic structures as proved using the mathematical tool of high-order homogenization in \cite{ysb2013}.

\section{Concluding remarks}
In conclusion, we have studied numerically the EM scattering properties of a cylindrical lens. Coordinates transformation can be used to realize a superscatterer with negatively refracting heterogeneous bianisotropic media, wherein a core with PEC boundary acts like a magnified PEC with radius $r_*$. Moreover, if the core is filled with certain bianisotropic media, a cloaking effect can be observed for a set of line dipoles lying at a specific distance from the shell, which can be attributed to the anomalous resonances of such kind of complementary media. Similarly, we explore the external cloaking effect in bianisotropic lenses; at low frequencies, an ostrich effect induced by the external cloaking can be observed with both the cylindrical lens and slab lens. Finally, it is possible to cloak finite size objects made of negatively refracting bianisotropic material at finite frequencies with slab and cylindrical lenses having a hole of same shape as the object, and opposite refractive index.
\end{document}